\documentclass[aps,prd,10pt,onecolumn,nofootinbib,longbibliography]{revtex4-2}
\usepackage{array}
\usepackage[english]{babel}
\usepackage{bm}

\usepackage{amsmath}
\usepackage{graphicx}
\usepackage{placeins}

\usepackage{amssymb}
\usepackage[caption=false]{subfig}

\usepackage{tikz}
\usetikzlibrary{calc,arrows.meta,decorations.pathmorphing,patterns}

\usepackage{siunitx}
\providecommand{\qty}{\SI}
\providecommand{\qtyrange}{\SIrange}
\providecommand{\unit}{\si}
\usepackage[colorlinks=true, allcolors=blue]{hyperref}
\usepackage{cleveref}
\crefname{equation}{Eq.}{Eqs.}
\Crefname{equation}{Equation}{Equations}
\crefname{figure}{Fig.}{Figs.}
\Crefname{figure}{Figure}{Figures}
\crefname{table}{Table}{Tables}
\Crefname{table}{Table}{Tables}

\newcommand{\real}[1]{\operatorname{Re}\left[#1\right]}
\newcommand{\imag}[1]{\operatorname{Im}\left[#1\right]}

\newcommand{\tc}[1]{#1^\dagger}

\newcommand{\sbsp}[3]{#1_{#2}^{#3}}
\newcommand{\vc}[1]{\bm{#1}}
\newcommand{\mt}[1]{\bm{#1}}

\newcommand{\diag}[1]{\operatorname{diag}\left[#1\right]}
\newcommand{\asqr}[1]{\left| #1 \right|^2}
\newcommand{\dsqr}[1]{\left| #1 \right|^{\circ 2}}

\newcommand{\nmsum}{\sum_n^\text{modes}}

\newcommand{\om}{(\omega)}
\newcommand{\omn}{(\omega_n)}

\newcommand{\kBT}{k_B T}
\newcommand{\kBTj}{k_B T_j}

\newcommand{\modal}[1]{\bar{#1}}
\newcommand{\kn}{\bar{k}_n}
\newcommand{\mn}{\bar{m}_n}
\newcommand{\cn}{\bar{c}_n}
\newcommand{\dn}{\bar{d}_n}
\newcommand{\phin}{\bar{\phi}_n}

\newcommand{\vx}{\vc{x}}
\newcommand{\vf}{\vc{f}}
\newcommand{\vy}{\vc{y}}

\newcommand{\vpsi}{\bm{\psi}}
\newcommand{\ve}{\vc{e}}

\newcommand{\mM}{\mt{M}}
\newcommand{\mK}{\mt{K}}
\newcommand{\mKe}{\mt{K}_e}
\newcommand{\mC}{\vc{C}}
\newcommand{\mD}{\mt{D}}
\newcommand{\mE}{\mt{E}}

\newcommand{\CAjdef}{\vpsi_A^T \mC \om \ve_j}
\newcommand{\CAj}{C_A^j}
\newcommand{\CAn}{C_{A,n}}
\newcommand{\CAnj}{C_{A,n}^j}

\newcommand{\CRj}{C_R^j}
\newcommand{\Cj}{C_j}

\newcommand{\GAj}{G_A^j}
\newcommand{\Gj}{G_j}

\newcommand{\anj}{\alpha_n^j}

\newcommand{\mGamma}{\mt{\Gamma}}

\newcommand{\Wdiss}{W_{\rm diss}}
\newcommand{\Umax}{U_{\rm max}}
\newcommand{\Umaxn}{U_{{\rm max},n}}
\newcommand{\UmaxA}{U_{{\rm max},A}}

\newcommand{\Sx}{S_x}
\newcommand{\Sf}{S_f}
\newcommand{\Svx}{S_{\vc{x}}}
\newcommand{\Svf}{S_{\vc{f}}}
\newcommand{\SxA}{S_{x_A}}
\newcommand{\SxR}{S_{x_R}}
\newcommand{\Sfj}{S_f^j}

\ifdefined\pdfminorversion\pdfminorversion=7\fi
\begin{document}
\raggedbottom
\title{A source-based decomposition of thermal noise for mechanical system design}
\author{Vermeer Bonhomme}
\author{Matthew Evans}
\affiliation{LIGO Laboratory, Massachusetts Institute of Technology, Cambridge, Massachusetts 02139, USA}

\begin{abstract}

Thermal noise limits the sensitivity of precision mechanical instruments. We present a thermal-noise formulation for lumped-element mechanical systems that writes the total displacement thermal noise at the readout as the sum of contributions from all dissipative elements. For each contribution, we separate the thermal force generated by the element from its transfer to the readout. This decomposition provides simple design rules for comparing contributions from elements in series and in parallel and identifying the dominant noise source. It applies to frequency-dependent dissipation and to elements at different steady temperatures under local thermal equilibrium. We use it to identify the frequency regimes in which parallel and series dissipation dilution reduce displacement thermal noise. We then use examples of a pendulum on a lossy support, a rotational flexure, and rigidly mounted lunar optics to show how support noise can dominate even when the support has little effect on the system dynamics. Finally, a feedback-controlled suspension illustrates how sensing and actuation noise can be included in the same displacement-noise budget.

\end{abstract}
\maketitle

\section{Introduction}

Thermal noise limits the sensitivity of many precision mechanical measurements, including gravitational-wave detectors \cite{instrO3paper,Gonzalez1994,Huang1998,Gretarsson2000,CEHS}, micro- and nanomechanical sensors \cite{Djuric2000,ButtJaschke1995,Bellon2008}, and ultrastable optical reference cavities used in precision frequency metrology and optical atomic clocks \cite{Numata2004}. More generally, it is an important limitation in low-noise force, displacement, acceleration, and frequency measurements where mechanical dissipation couples to the measured degree of freedom.

In this paper, we develop a systematic method for identifying
the dominant thermal-noise sources in mechanical systems. We derive
simple rules for comparing noise contributions without requiring
detailed finite-element analysis. The method is particularly useful for systems with multiple dissipative
elements, such as multistage suspensions, where thermal noise generated
at different stages can be filtered differently before reaching the
readout~\cite{Gonzalez1994,Huang1998,Gretarsson2000,CEHS}.
It also accommodates components at different temperatures and
frequency-dependent dissipation mechanisms, including viscoelastic
materials and dissipation in resistively shunted piezoelectric elements,
as used in acoustic mode dampers~\cite{Gras2015AMD,Biscans2019AMD}. In particular, we show that supporting structures with negligible influence on the system dynamics can nevertheless dominate the readout noise.

Several equivalent formulations of the fluctuation--dissipation theorem
are commonly used to calculate the total thermal noise at the readout,
but the total spectrum alone gives little insight into where the
noise is generated or how it reaches the readout.
Levin's virtual-force method obtains the thermal noise from the power
dissipated under a force with the spatial profile of the
readout~\cite{levin1998}. It is particularly well suited to detailed
finite-element models at later design stages, when the component
geometry and material properties are known in enough detail to
calculate how each part deforms and how much power it dissipates
under the applied force~\cite{BraginskyLevinVyatchanin1999,Harry2002,Nakagawa2002,Grass2026}. Saulson's formulation instead expresses the thermal noise directly in terms of the mechanical admittance of the complete system \cite{Saulson1990jc}, without directly exposing how that noise is distributed among the individual dissipative elements~\cite{Gonzalez1994,Ohishi2000,Ruggi2025}.
More recent publications have explored the separation of thermal-noise contributions from individual elements
in branched oscillator networks \cite{Puppo2010} and in temperature-weighted matrix descriptions of coupled
mechanical systems \cite{Komori2018}.

Building on these approaches, we organize the thermal-force source
description into a systematic form for lumped-element mechanical systems.
The framework makes explicit both the noise generated by each dissipative
element and its transmission to the readout.
Because the method separates the generation of noise from its propagation through the system, the same framework can also accommodate nonthermal noise sources and additional dynamical couplings, including feedback-control forces and optomechanical interactions, while retaining a clear attribution of how each source contributes to the final readout. 

We first clarify how parallel and series dissipation dilution reduce
thermal noise in different frequency regimes. We then compare the
competing thermal-noise sources in a suspended mass on a lossy
support, a rotational flexure representative of six-dimensional
inertial sensors~\cite{Smetana2025,Saffarieh2026}, and rigidly
mounted optics for a lunar interferometer such as the Laser Interferometer
Lunar Antenna (LILA)~\cite{Creighton2025LILA}. Finally, we use a feedback-controlled
suspension to show how sensing and actuation noise enter the
displacement-noise budget alongside thermal noise.

\section{Main result}
%\section{Source-Based Thermal-Noise Formulation}
Consider a mechanical system composed of masses connected by elastic elements, with the displacement of some part of the system serving as the readout. Each element with mechanical loss generates a fluctuating force between the points it connects, producing motion throughout the system. The displacement noise observed at the readout therefore depends on both the magnitude of these fluctuating forces and the mechanical response of the system to each of them.

The contribution of an individual element can be evaluated by imagining
a force applied across that element and calculating the resulting
displacement at the readout. This response is described by a dynamic
compliance, which includes the effects of all the masses, stiffnesses,
and mechanical resonances in the system.
Assuming that the thermal forces generated by different elements
are uncorrelated, their contributions to the readout
displacement-noise power spectral density (PSD) add independently,
giving
\begin{equation}
 \SxR\om=\sum_j\asqr{\Cj\om}\Sfj\om,
 \label{eq:Sx A}
\end{equation}
where $\Sfj$ is the thermal-force PSD of element $j$, $C_j$
is the dynamic compliance from a force applied across that element
to the readout, and $\omega$ is the angular frequency.
Appendix~\ref{sec:derivation} derives this
expression from Levin's method.
Assuming locally equilibrated baths with approximately steady
temperatures $T_j$~\cite{Komori2018},
the thermal-force spectrum of element $j$ is
\begin{equation}
    \Sfj\om
    =
    \frac{4\kBTj}{\omega}
    \imag{k_j\om},
    \label{eq:SFj}
\end{equation}
where $k_j(\omega)$ is the element's generalized stiffness.
This expression applies to viscous damping, structural damping,
and other passive mechanical losses represented by its imaginary part. 
%A similar analysis is well established for electrical systems,
%as discussed in the feedback-controlled suspension example in \cref{sec:electrical_analogy}.

\Cref{eq:Sx A} can be rewritten to separate out the dynamic compliance of the system at the readout, $C_R\om$, from the transfer of force from elements of the system to the readout, $\Gj\om$, as 
\begin{equation}
\label{eq:Sx B}
    \SxR\om
    =
    \asqr{C_R\om}
    \sum_j
    \asqr{\Gj\om}
    \Sfj\om,
\end{equation}
where $C_R$ is the compliance from a generalized force applied at
the readout to its displacement, and $G_j=C_j/C_R$ is the
equivalent force-transfer factor for element $j$.

Important corollaries to \cref{eq:Sx B}, derived in
Appendix~\ref{sec:Saulson}, provide simple design rules
by expressing the relative thermal-noise contributions of elements
in parallel and in series.  Elements in parallel have $G_1\om = G_2\om$ such that
\begin{equation}
\frac{\SxR^1\om}{\SxR^2\om}
=
\frac{S_f^1\om}{S_f^2\om}
= \frac{T_1}{T_2}
\frac{\imag{k_1\om}}{\imag{k_2\om}}
=
\frac{T_1}{T_2}\frac{\phi_1\om k_1'\om}{\phi_2\om k'_2\om },
\label{eq:parallel thermal noise comparison}
\end{equation}
where $k_j'\equiv \real{k_j}$ and
the loss angle $\phi_j \equiv \imag{k_j} / \real{k_j}$.
Similarly, for elastic elements in series with massless, unbranched junctions,
\begin{equation}
\frac{\SxR^1\om}{\SxR^2\om}
=
\frac{|G_1\om|^2 S_f^1\om}{|G_2\om|^2 S_f^2\om}
=
\frac{T_1}{T_2}\frac{\imag{k_1\om}/{|k_1\om|^2}}{\imag{k_2\om}/{|k_2\om|^2}}
\simeq
\frac{T_1}{T_2}\frac{\phi_1\om / k_1'\om}{\phi_2\om/{k_2'\om}},
\label{eq:series ratio}
\end{equation}
with the approximation valid for small loss angles $\phi_1\om, \phi_2\om \ll 1$ such that $k_j' \simeq |k_j|$. 

To calculate the compliances and force-transfer factors for a given mechanical system,
we represent the system by nodes connected by stiffness elements. The vectors
$\vx=(x_1,\ldots,x_N)^T$ and $\vf$
contain the displacements of the nodes and the forces applied to them,
respectively. The frequency-domain equations of motion give
\begin{equation}
 \vx(\omega)=\mC(\omega)\vf(\omega),\qquad
 \mC(\omega)=[\mK(\omega)-\omega^2\mM]^{-1},
 \label{eq:main_nodal_response}
\end{equation}
where $\mM$ and $\mK$ are the nodal mass and complex stiffness matrices.
The dynamic compliance matrix $\mC$ therefore includes the effects
of the masses, stiffnesses, and mechanical resonances.

The weighting vector $\vpsi_R$ projects the nodal displacements onto
the readout, while the connection vector $\ve_j$ gives the elongation
of element $j$,
\begin{equation}
 x_R=\vpsi_R^T\vx,\qquad y_j=\ve_j^T\vx.
\end{equation}
For example, if element $j$ connects nodes $a$ and $b$ with
$y_j=x_b-x_a$, then $\ve_j$ has entries $-1$ at node $a$,
$+1$ at node $b$, and zero elsewhere. A force $f_j$ applied across
that element produces the nodal force vector $\vf=\ve_j f_j$.
With this force alone applied, the readout response is
\begin{equation}
 x_R(\omega)=C_j(\omega)f_j(\omega),\qquad
 C_j(\omega)\equiv\vpsi_R^T\mC(\omega)\ve_j.
 \label{eq:main_source_compliance}
\end{equation}
The readout self-compliance is $C_R=\vpsi_R^T\mC\vpsi_R$.
Appendix~\ref{sec:notation} gives further details of this construction.

To briefly distinguish these results from recent related work, we note that \cref{eq:Sx A} is the element-wise form of the temperature-weighted matrix
expressions in Eqs.~(6)--(7) of Komori et al.~\cite{Komori2018}.
While formally equivalent, 
unlike the matrix formulation
in \cite{Komori2018} our \cref{eq:Sx A} makes explicit both the
individual noise source and its transmission path to the readout.
The result presented in Eq.~(12) of Puppo \cite{Puppo2010},
on the other hand, clearly separates noise generation from transmission, but the analysis is limited to branched oscillator configurations and does not allow for general lumped element models.

We apply these comparisons to several high-sensitivity instruments in
\cref{sec:Applications} and conclude in \cref{sec:conclusion}. The appendices detail the formalism and derivation of the thermal-noise transfer functions, derive the source-based formulation from Levin's method, establish its connection to Saulson's formulation, and extend the framework to continuous systems, including a detailed treatment of modal analysis.

\section{Applications}
\label{sec:Applications}

\subsection{Dissipation dilution}

\begin{figure}[tbp]
    \captionsetup[subfloat]{font={normalfont,normalsize,bf},labelfont=bf,textfont=bf}

    \centering

\parbox[t]{0.48\linewidth}{\centering \subfloat[Parallel configuration.\label{fig:loss_parallel}]{\makebox[\linewidth]{}}}%
\hfill%
\parbox[t]{0.48\linewidth}{\centering \subfloat[Series configuration.\label{fig:loss_series}]{\makebox[\linewidth]{}}}\par\vspace{2mm}
$\vcenter{\hbox{\resizebox{0.48\linewidth}{!}{%
\begin{tikzpicture}[
            line width=0.9pt,
            font=\small,
            spring/.style={
                decorate,
                decoration={
                    zigzag,
                    segment length=5pt,
                    amplitude=3pt
                }
            },
            wall/.style={
                pattern=north east lines,
                draw,
                minimum width=0.12cm,
                minimum height=2.4cm
            },
            mass/.style={
                draw,
                rounded corners=1pt,
                minimum width=1.15cm,
                minimum height=0.95cm,
                fill=gray!12
            }
        ]

        \node[wall] (wall) at (0,0) {};

        \coordinate (topL) at (0.1,0.55);
        \coordinate (botL) at (0.1,-0.55);

        \draw (wall.east |- topL) -- (0.65,0.55);

        \draw[red!75!black,spring]
            (0.65,0.55) -- (2.65,0.55);

        \draw (2.65,0.55) -- (4.25,0.55);

        \draw (wall.east |- botL) -- (0.65,-0.55);

        \draw[spring]
            (0.65,-0.55) -- (2.65,-0.55);

        \draw (2.65,-0.55) -- (4.25,-0.55);

        \draw (4.25,0.55) -- (4.25,-0.55);

        \fill (4.25,0.55) circle (1.5pt);
        \fill (4.25,-0.55) circle (1.5pt);

        \node[mass] (M) at (5.00,0) {$m$};

        \draw (4.25,0) -- (M.west);

        \draw[blue!70!black,line width=1.1pt]
            (M.east) -- ++(.75,0);

        \fill[blue!70!black]
            ([xshift=.75cm]M.east)
            -- ++(-0.16cm,0.09cm)
            -- ++(0,-0.18cm)
            -- cycle;

        \node[black,right=4pt]
            at ([xshift=0.7cm]M.east) {$x_{\scriptscriptstyle R}$};

        \node[red!75!black,above=5pt,align=center]
            at (1.65,0.55)
            {$k'_\ell(1+i\phi_\ell)$};

        \node[below=6pt,align=center]
            at (1.65,-0.55)
            {$k_0'$};

        \end{tikzpicture}%
}}}$%
\hfill%
$\vcenter{\hbox{\resizebox{0.48\linewidth}{!}{%
\begin{tikzpicture}[
            line width=0.9pt,
            font=\small,
            spring/.style={
                decorate,
                decoration={
                    zigzag,
                    segment length=5pt,
                    amplitude=3pt
                }
            },
            wall/.style={
                pattern=north east lines,
                draw,
                minimum width=0.12cm,
                minimum height=1.4cm
            },
            mass/.style={
                draw,
                rounded corners=1pt,
                minimum width=1.15cm,
                minimum height=0.85cm,
                fill=gray!12
            }
        ]

        \node[wall] (wall) at (0,0) {};

        \draw (wall.east) -- (0.55,0);

        \draw[red!75!black,spring]
            (0.55,0) -- (2.55,0);

        \fill (2.55,0) circle (1.5pt);

        \draw (2.55,0) -- (3.15,0);

        \draw[spring]
            (3.15,0) -- (4.85,0);

        \node[mass] (M) at (5.55,0) {$m$};

        \draw (4.85,0) -- (M.west);

        \draw[blue!70!black,line width=1.1pt]
            (M.east) -- ++(.75,0);

        \fill[blue!70!black]
            ([xshift=0.75cm]M.east)
            -- ++(-0.16cm,0.09cm)
            -- ++(0,-0.18cm)
            -- cycle;

        \node[black,right=4pt]
            at ([xshift=0.7cm]M.east) {$x_{\scriptscriptstyle R}$};

        \node[red!75!black,above=7pt]
            at (1.55,0)
            {$k_\ell'(1+i\phi_\ell)$};

        \node[above=7pt]
            at (4.00,0)
            {$k_0'$};

        \end{tikzpicture}%
}}}$\par
\caption{
        Simple spring--mass systems illustrating mechanical-loss dilution.
        \textbf{(a)} A lossy spring with complex stiffness
        $k'_\ell(1+i\phi_\ell)$ acts in parallel with a lossless spring
        of stiffness $k'_0$.
        \textbf{(b)} A lossy spring with complex stiffness
        $k'_\ell(1+i\phi_\ell)$ acts in series with a lossless spring
        of stiffness $k'_0$.
        In both cases, $x_R$ denotes the displacement of the mass $m$.
    }

    \label{fig:loss_dilution}

\end{figure}
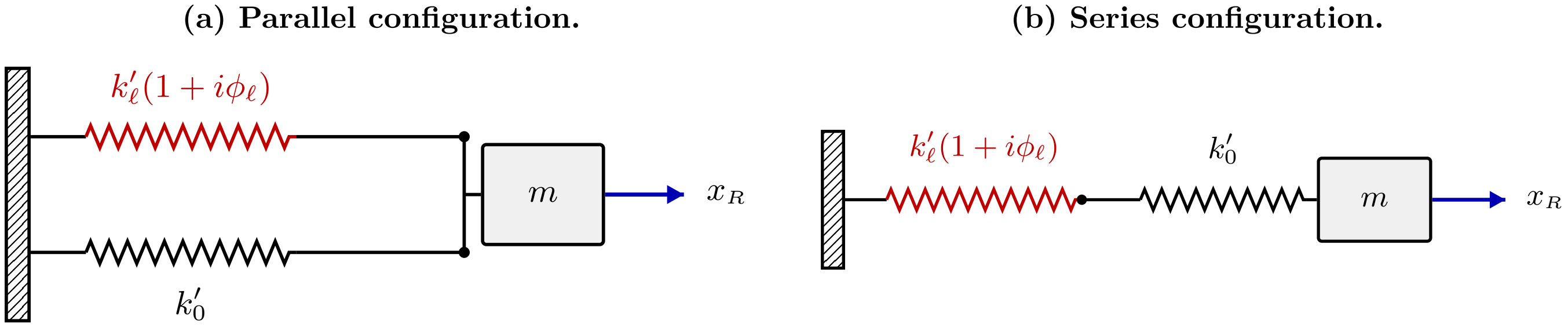

\textcolor{black}{Dissipation dilution refers to the reduction of the effective loss angle of a mechanical system when part of its restoring energy is stored in a lossless element, such as the gravitational restoring force of a pendulum.} This principle is commonly described and used to reduce thermal noise in mechanical systems. \textcolor{black}{The description in terms of individual thermal-force sources makes clear, however, that a reduced effective loss angle does not necessarily imply a corresponding reduction of the displacement thermal noise at all frequencies.} The two elementary configurations considered below, with a lossless stiffness placed either in parallel or in series with a lossy stiffness, are illustrated in Fig.~\ref{fig:loss_dilution}. 
%Thecomparisons hold $m$, temperature, and the lossy spring properties fixed while changing the lossless stiffness.

\subsubsection{Parallel dilution}
\label{sec:parallel dilution}

Consider a mass $m$ connected to a lossy spring $k_\ell(\omega)=k'_\ell\om(1+i\phi_\ell\om)$ and to an additional lossless spring $k_0'\om$ placed in parallel. Denoting the displacement of the mass by $x_R$, the dynamic compliance is
\begin{equation}
    C_R(\omega)
    =
    \frac{1}
    {-m\omega^2+k_0'\om+k_\ell(\omega)}.
\end{equation}
For frequency-independent stiffnesses, the undamped natural
angular frequency is
\begin{equation}
    \omega_0^2=\frac{k_0'+k'_{\ell}}{m}.
\end{equation}

Since the lossless spring produces no thermal-force noise, the displacement thermal noise is
\begin{equation}
    S_{x_R}(\omega)
    =
    \left|
    \frac{1}
    {-m\omega^2+k_0'\om+k_\ell(\omega)}
    \right|^2
    \frac{4k_B T}{\omega}
    \imag{k_\ell(\omega)}.
\end{equation}
In the quasistatic regime, the inertial term can be neglected.
For frequency-independent elastic stiffnesses and small loss,
this corresponds to frequencies well below resonance,
$\omega\ll\omega_0$, giving
\begin{equation}
    S_{x_R}(\omega)
    \simeq
    \left|
    \frac{1}
    {k_0'(\omega)+k_\ell(\omega)}
    \right|^2
    \frac{4k_B T}{\omega}
    \imag{k_\ell(\omega)}.
\end{equation}
Relative to the system containing only the lossy spring, the \textcolor{black}{displacement thermal-noise PSD is therefore reduced by the factor}
\begin{equation}
    \left|
    \frac
    {k_0'(\omega)+k_\ell(\omega)}{k_\ell(\omega)}
    \right|^2.
\end{equation}

\begin{figure}[tbp]
  \centering
  \includegraphics[width=0.72\linewidth]{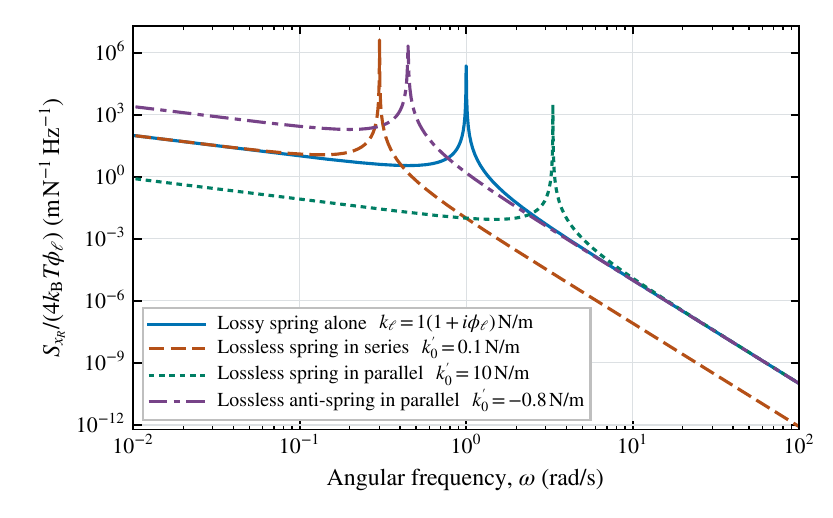}
  \caption{Thermal displacement-noise PSD for a mass $m=\qty{1}{\kilogram}$ connected to a lossy spring alone, with an additional lossless spring in parallel, with a lossless spring in series, and with a lossless anti-spring in parallel. Parallel dilution reduces the displacement noise below resonance but leaves the far-above-resonance noise unchanged, whereas series dilution produces the complementary behavior. The lossy spring has stiffness $k_\ell'=\qty{1}{\newton\per\metre}$ and loss angle $\phi_\ell=\num{1e-3}$. The added spring stiffness is \qty{0.1}{\newton\per\metre} for the series case, \qty{10}{\newton\per\metre} for the parallel case, and \qty{-0.8}{\newton\per\metre} for the anti-spring case. The PSD is normalized by $4k_BT\phi_\ell$.}
  \label{fig:thermal_noise_dilution}
\end{figure}

\textcolor{black}{This reduction is illustrated in Fig.~\ref{fig:thermal_noise_dilution}, where the parallel configuration has a lower noise level than the original system at frequencies well below both resonances.}

In experiments using suspended optics, as LIGO does~\cite{instrO3paper},
however, the relevant observation band is generally above the
suspension resonance. We therefore consider the inertia-dominated
regime, where the restoring forces can be neglected. For
frequency-independent elastic stiffnesses and small loss, this
corresponds to $\omega\gg\omega_0$, and the thermal noise is approximately
\begin{equation}
    S_{x_R}(\omega)
    \simeq
    \frac{4k_B T}{m^2\omega^5}
    \imag{k_\ell(\omega)},
    \label{eq:pdilu}
\end{equation}
\textcolor{black}{which is identical to the above-resonance thermal noise of the lossy spring alone. Thus, adding a lossless spring in parallel does not reduce the displacement thermal noise above resonance. This behavior is visible in Fig.~\ref{fig:thermal_noise_dilution}, where the parallel and original spectra converge at high frequency.}

\textcolor{black}{
This expression can equivalently be written in the classical form used by Gonz\'alez~\cite{Gonzalez2000},
}
\begin{equation}
    S_{x_R}(\omega)
    \simeq
    \frac{4k_B T\,\omega_0^2\phi_{\parallel}}
    {m\omega^5},
\end{equation}
\textcolor{black}{
where the effective loss angle is reduced by the dilution factor
}
\begin{equation}
    \phi_{\parallel}
    \simeq
    \frac{\phi_\ell}{D_{\parallel}},
    \qquad
    D_{\parallel}
    \equiv
    \frac{U_0+U_{\rm loss}}{U_{\rm loss}}
    =
    \frac{k_0'+k_\ell'}{k_\ell'},
\end{equation}
\textcolor{black}{
where $U_0$ and $U_{\rm loss}$ are the elastic energies stored in the lossless and lossy springs, respectively, giving the usual stored-energy expression for dissipation dilution~\cite{Gonzalez1994,Saulson1990jc}.
}At first sight, the reduction of $\phi_{\parallel}$ may suggest a corresponding reduction of the above-resonance thermal noise. However, this description replaces the two parallel springs by an equivalent spring. The addition of the lossless stiffness reduces the effective loss angle while simultaneously increasing the resonance frequency, such that the two changes cancel in the combination $\omega_0^2\phi_{\parallel}$ entering the above-resonance thermal noise.

\textcolor{black}{
In the thermal-force-source description used here, the fact that the thermal noise is reduced below resonance but not above resonance follows immediately. We consider the thermal force generated by the lossy spring and examine how the system responds to this same force in the two frequency regimes.
}\textcolor{black}{
Below resonance, the motion is quasistatic and the additional lossless stiffness reduces the displacement produced by the unchanged thermal force. Far above resonance, the response is inertia dominated, the restoring forces of the lossless spring can be neglected, and the same thermal force produces the same displacement whether or not the lossless parallel stiffness is present. Thus, although the system exhibits dissipation dilution, the far-above-resonance displacement thermal noise is not reduced.
}

In particular, the gravitational restoring force does not enter the above-resonance thermal noise explicitly. For a fixed thermal-force spectrum, increasing the suspended mass reduces the displacement noise because the larger inertia produces less displacement under the same force, rather than because more energy is stored in the gravitational spring.

\medskip
\paragraph{Lossless anti-spring in parallel}

Lossless gravitational anti-springs are used in accelerometers and
other inertial sensors to reduce the effective stiffness of the
system. This increases the displacement
response to an applied acceleration, making the sensing noise smaller
relative to the signal. It also lowers the resonance frequency, extending the frequency range over which the proof mass serves as an inertial reference. Examples are the tilt modes of six-axis inertial sensors
developed for gravitational-wave observatories~\cite{Smetana2025,Saffarieh2026}.

It may be tempting to associate the lower effective stiffness, or
lower resonance frequency, with a corresponding reduction of the
thermal noise. Assigning the loss angle of the lossy spring to the
reduced total stiffness would, however, effectively assign negative
dissipation to the lossless anti-spring. 
%In the following comparison,the mass and lossy-element stiffness are held fixed.

The thermal-noise behavior of spring--anti-spring systems has been
derived explicitly for Roberts linkages and flexure-supported
beam-balance tiltmeters considered for gravitational-wave
instrumentation, where the post-resonance thermal noise is found
not to be reduced~\cite{HarmsMowLowry2018}. In the thermal-force-source description, this result follows directly as the $k_0'<0$ case of the parallel-dilution result above. Below resonance, the reduced effective stiffness increases the displacement produced by the unchanged thermal force, whereas far above resonance the response is inertia dominated and the thermal-noise level is unchanged. Thus, the limited post-resonance thermal-noise reduction of an anti-spring does not require a separate treatment, but follows directly from its parallel topology. Figure~\ref{fig:thermal_noise_dilution} illustrates this behavior.

\subsubsection{Series dilution}
\label{sec:series dilution}

Consider a mass $m$ connected to a lossy spring $k_\ell(\omega)$ in series with a lossless spring $k_0'(\omega)$. Denoting the displacement of the mass by $x_R$, the dynamic compliance is
\begin{equation}
    C_R(\omega)
    =
    \frac{1}
    {-m\omega^2+
    \dfrac{k_\ell(\omega)k_0'(\omega)}
    {k_\ell(\omega)+k_0'(\omega)}}.
\end{equation}
For frequency-independent stiffnesses, the undamped natural
angular frequency is
\begin{equation}
    \omega_0^2
    =
    \frac{1}{m}
    \frac{k_\ell'k_0'}
    {k_\ell'+k_0'}.
\end{equation}

Since the lossless spring generates no thermal force, the displacement thermal noise is
\begin{equation}
    S_{x_R}(\omega)
    =
    \left|
    \frac{1}
    {-m\omega^2+
    \dfrac{k_\ell(\omega)k_0'(\omega)}
    {k_\ell(\omega)+k_0'(\omega)}}
    \right|^2
    \left|
    \frac{k_0'(\omega)}
    {k_\ell(\omega)+k_0'(\omega)}
    \right|^2
    \frac{4k_B T}{\omega}
    \imag{k_\ell(\omega)}.
\end{equation}

In the quasistatic regime ($\omega\ll\omega_0$), the thermal noise is

\begin{equation}
    S_{x_R}(\omega)
    \simeq
    \left|
    \frac{1}{k_\ell(\omega)}
    \right|^2
    \frac{4k_B T}{\omega}
    \imag{k_\ell(\omega)}.
\end{equation}
The additional lossless spring therefore does not reduce the displacement thermal noise below resonance. 

In the inertia-dominated regime ($\omega\gg\omega_0$), the thermal noise is
\begin{equation}
    S_{x_R}(\omega)
    \simeq
    \frac{4k_B T}{m^2\omega^5}
    \left|
    \frac{k_0'(\omega)}
    {k_\ell(\omega)+k_0'(\omega)}
    \right|^2
    \imag{k_\ell(\omega)}.
    \label{eq:sdilu}
\end{equation}
Relative to the system containing only the lossy spring, the displacement thermal-noise PSD is therefore reduced by the factor
\begin{equation}
    \left|
    \frac{k_\ell(\omega)+k_0'(\omega)}
    {k_0'(\omega)}
    \right|^2.
\end{equation}
Efficient above-resonance thermal-noise reduction is thus obtained by placing the lossy spring in series with a much softer lossless spring. The corresponding high-frequency suppression is shown by the lower series-configuration spectrum in Fig.~\ref{fig:thermal_noise_dilution}.

This expression can equivalently be written in terms of the effective properties of the equivalent spring,
\begin{equation}
    S_{x_R}(\omega)
    \simeq
    \frac{4k_B T\,\omega_0^2\phi_+}
    {m\omega^5}.
\end{equation}
The effective loss angle is reduced by the dilution factor
\begin{equation}
    \phi_+
    \simeq
    \frac{\phi_\ell}{D_{\rm +}},
    \qquad
    D_{\rm +}
    \equiv
    \frac{U_0+U_{\rm loss}}{U_{\rm loss}}
    =
    \frac{k_\ell'+k_0'}{k_0'}.
\end{equation}
In contrast with the parallel configuration, the reduction of the effective loss angle is accompanied by a reduction of the resonance frequency. The two effects therefore reinforce one another in the above-resonance thermal noise rather than cancel.

In the thermal-force-source description, the behavior of the series
configuration follows from how the thermal force generated by the
lossy spring is transmitted to the mass. Below resonance, the response is quasistatic, and the displacement
produced by the thermal force is the same as for the lossy spring alone. Far above resonance,
the response of the mass is inertia dominated, but only a fraction (or force-transfer factor) $G_\ell(\omega)=k_0'(\omega)/[k_\ell(\omega)+k_0'(\omega)]$ 
of the thermal force generated by the lossy spring is transmitted to the mass. The lossless series spring therefore reduces the above-resonance displacement thermal noise.

\subsection{Competing thermal-noise sources}
\label{sec:competing_thermal-noise_sources}
Lossy elements in series can contribute very differently to the total
readout noise. By \cref{eq:series ratio}, the relevant comparison at
equal temperatures and small loss angles is the loss angle divided
by the elastic stiffness, $\phi_j/k_j'$. A stiff support can therefore
dominate the readout noise even when its compliance has little effect
on the system dynamics. We compare these contributions for a suspended
mass, a rotational flexure, and rigidly mounted optics. In all three
examples, the elastic stiffnesses and structural loss angles are taken
to be frequency independent.

\subsubsection{Suspended mass on a lossy support}

We now consider a suspension with both intrinsic suspension loss and loss in its supporting structure. The physical configuration and its equivalent lumped-element model are shown in Fig.~\ref{fig:suspended_mass_lossy_support}. We compare the support contribution directly with the thermal noise generated by the suspension itself.

\usetikzlibrary{
    arrows.meta,
    decorations.pathmorphing,
    patterns,
    angles,
    quotes
}

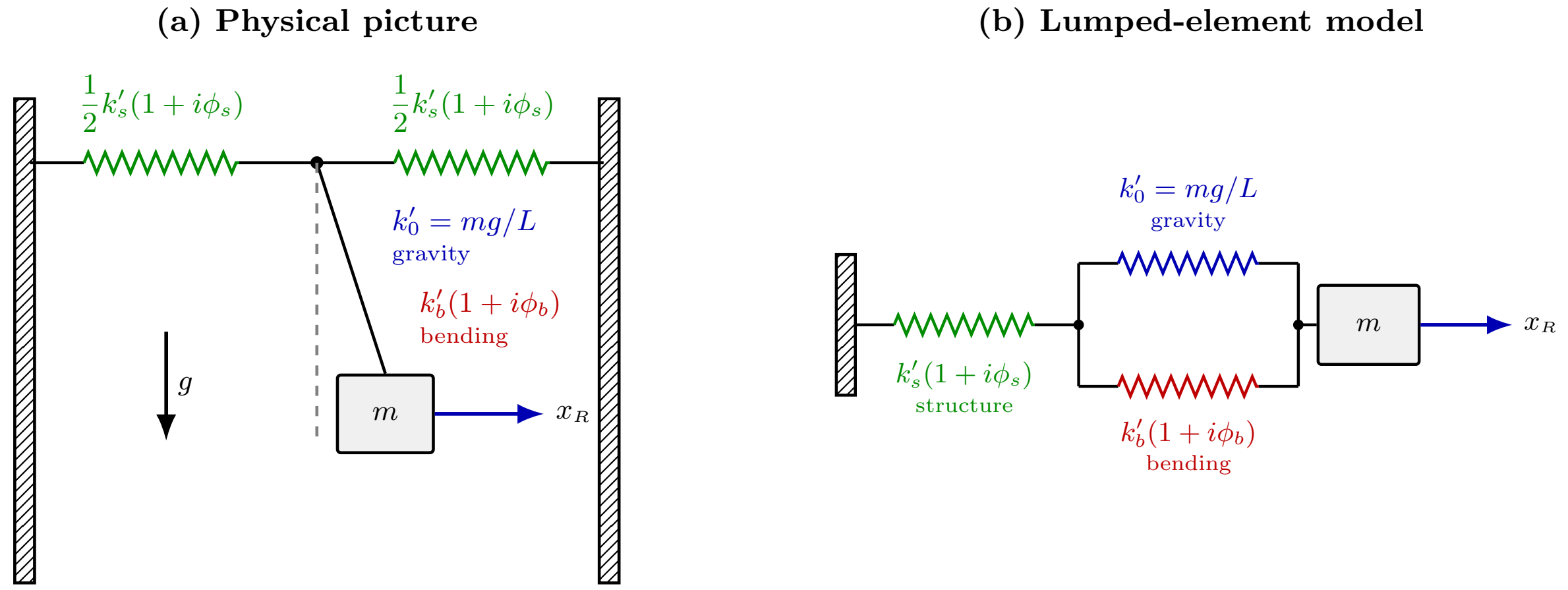
\begin{figure}[ht]
\centering
\parbox[t]{0.34\linewidth}{\centering \bfseries\strut (a) Physical picture}%
\hspace{.12\linewidth}%
\parbox[t]{0.41\linewidth}{\centering \bfseries\strut (b) Lumped-element model}\par\vspace{2mm}
$\vcenter{\hbox{\resizebox{0.34\linewidth}{!}{%
\begin{tikzpicture}[
    line width=0.9pt,
    >=Latex,
    font=\small,
    scale=0.95,
    transform shape,
    spring/.style={
        decorate,
        decoration={
            zigzag,
            segment length=5pt,
            amplitude=3pt
        }
    },
    wall/.style={
        pattern=north east lines,
        draw,
        minimum width=0.12cm,
        minimum height=5.3cm
    },
    mass/.style={
        draw,
        rounded corners=1pt,
        minimum width=1.05cm,
        minimum height=0.85cm,
        fill=gray!12
    }
]

\node[wall] (WL) at (0,-0.45) {};
\node[wall] (WR) at (6.4,-0.45) {};

\coordinate (WLattach) at (0.06,1.5);
\coordinate (WRattach) at (6.34,1.5);

\draw (WLattach) -- (0.65,1.5);

\draw[
    green!55!black,
    spring
]
    (0.65,1.5) -- (2.35,1.5);

\draw (2.35,1.5) -- (3.2,1.5);

\fill (3.2,1.5) circle (2pt);

\draw (3.2,1.5) -- (4.05,1.5);

\draw[
    green!55!black,
    spring
]
    (4.05,1.5) -- (5.75,1.5);

\draw (5.75,1.5) -- (WRattach);

\node[
    green!55!black,
    above=6pt
] at (1.50,1.5)
{
    $\dfrac{1}{2}k'_s(1+i\phi_s)$
};

\node[
    green!55!black,
    above=6pt
] at (4.90,1.5)
{
    $\dfrac{1}{2}k'_s(1+i\phi_s)$
};

\coordinate (P) at (3.2,1.5);
\coordinate (V) at (3.2,-1.55);

\node[mass] (M) at (3.95,-1.25) {$m$};

\draw (P) -- (M.north);
\draw[dashed,gray] (P) -- (V);

% Gravity direction in the physical view.
\ifdefined\DIFadd
    \colorlet{gravityarrow}{green!36!black}
\else
    \colorlet{gravityarrow}{black}
\fi
\draw[gravityarrow,->,line width=1.1pt]
    (1.55,-0.35) -- (1.55,-1.55)
    node[midway,right] {$g$};

\node[
    blue!70!black,
    anchor=west,
    align=left
] at (3.9,0.7)
{
    $k'_0=mg/L$\\[-1pt]
    \scriptsize gravity
};

\node[
    red!75!black,
    anchor=west,
    align=left
] at (4.20,-0.20)
{
    $k'_b(1+i\phi_b)$\\[-1pt]
    \scriptsize bending
};

\draw[
    blue!70!black,
    ->,
    line width=1.1pt
]
    (M.east) -- ++(1.2,0)
    node[right,black] {$x_{\scriptscriptstyle R}$};

\end{tikzpicture}%
}}}$%
\hspace{.12\linewidth}%
$\vcenter{\hbox{\resizebox{0.41\linewidth}{!}{%
\begin{tikzpicture}[
    line width=0.9pt,
    >=Latex,
    font=\small,
    scale=0.95,
    transform shape,
    spring/.style={
        decorate,
        decoration={
            zigzag,
            segment length=5pt,
            amplitude=3pt
        }
    },
    wall/.style={
        pattern=north east lines,
        draw,
        minimum width=0.12cm,
        minimum height=1.6cm
    },
    mass/.style={
        draw,
        rounded corners=1pt,
        minimum width=1.15cm,
        minimum height=0.9cm,
        fill=gray!12
    }
]

\node[wall] (wall) at (0,0) {};

\draw (wall.east) -- (0.55,0);

\draw[
    green!55!black,
    spring
]
    (0.55,0) -- (2.15,0);

\draw (2.15,0) -- (2.65,0);

\fill (2.65,0) circle (1.7pt);

\draw (2.65,0) -- (2.65,0.7);
\draw (2.65,0.7) -- (3.10,0.7);

\draw[
    blue!70!black,
    spring
]
    (3.10,0.7) -- (4.70,0.7);

\draw (4.70,0.7) -- (5.15,0.7);
\draw (5.15,0.7) -- (5.15,0);

\draw (2.65,0) -- (2.65,-0.7);
\draw (2.65,-0.7) -- (3.10,-0.7);

\draw[
    red!75!black,
    spring
]
    (3.10,-0.7) -- (4.70,-0.7);

\draw (4.70,-0.7) -- (5.15,-0.7);
\draw (5.15,-0.7) -- (5.15,0);

\fill (5.15,0) circle (1.7pt);

\node[mass] (M) at (5.95,0) {$m$};

\draw (5.15,0) -- (M.west);

\draw[
    blue!70!black,
    ->,
    line width=1.1pt
]
    (M.east) -- ++(1.05,0)
    node[right,black] {$x_{\scriptscriptstyle R}$};

\node[
    green!55!black,
    below=8pt,
    align=center
] at (1.35,0)
{
    $k'_s(1+i\phi_s)$\\[-1pt]
    \scriptsize structure
};

\node[
    blue!70!black,
    above=7pt,
    align=center
] at (3.90,0.7)
{
    $k'_0=mg/L$\\[-1pt]
    \scriptsize gravity
};

\node[
    red!75!black,
    below=7pt,
    align=center
] at (3.90,-0.7)
{
    $k'_b(1+i\phi_b)$\\[-1pt]
    \scriptsize bending
};

\end{tikzpicture}%
}}}$\par
\caption{
Suspended mass on a compliant supporting structure.
\textbf{(a)} Physical configuration of a pendulum of mass \(m\) suspended from a lossy supporting structure. The pendulum restoring stiffness contains a lossless gravitational contribution \(k_0'=mg/L\) in parallel with the lossy bending stiffness \(k_b'(1+i\phi_b)\), while the supporting structure has stiffness \(k_s'(1+i\phi_s)\).
\textbf{(b)} Equivalent lumped-element model showing the structural stiffness in series with the parallel gravitational and bending stiffnesses of the pendulum. The displacement \(x_R\) of the suspended mass is taken as the readout.
}
\label{fig:suspended_mass_lossy_support}
\end{figure}

The effective pendulum stiffness $k_p=k_0'+k_b$ combines the
lossless gravitational stiffness $k_0'$ with the lossy bending
stiffness $k_b=k_b'(1+i\phi_b)$ of the suspension wire or fiber.
For $k_b'\ll k_0'$, this gives
\begin{equation}
    k_p
    \simeq
    k_0'
    \left[
        1+i\frac{k_b'}{k_0'}\phi_b
    \right].
\end{equation}
The corresponding effective pendulum loss angle is therefore
\begin{equation}
    \phi_p
    \simeq
    \frac{k_b'}{k_0'}\phi_b.
\end{equation}
The supporting structure is represented by
\begin{equation}
    k_s=k_s'(1+i\phi_s).
\end{equation}

Since the pendulum stiffness and the structural stiffness are
in series, \cref{eq:series ratio} gives the condition for the
structural thermal noise to dominate at equal temperature,
\begin{equation}
    \frac{\phi_s}{k_s'}
    >
    \frac{\phi_p}{k_p'}
    \simeq
    \frac{k_b'\phi_b}{k_0'^2}.
\end{equation}
Because the suspension loss is strongly diluted by gravity, and
the pendulum may be suspended with low-loss materials
(e.g., fused silica, with $\phi_b\sim10^{-8}$~\cite{3Gsusp}),
the structural loss can easily dominate.

Taking the final stage of the Advanced LIGO suspension as an
example, its $\qty{40}{\kilogram}$ test mass, approximately
$\qty{0.6}{\metre}$ suspension length, and pendulum dilution
factor of about $90$~\cite{Aston2012} correspond, in this
effective model, to
\begin{equation}
    k_0'\simeq\qty{650}{\newton\per\metre},
    \qquad
    k_b'\simeq\qty{7.3}{\newton\per\metre}.
\end{equation}
Assuming a steel support with effective loss angle $\phi_s=10^{-3}$,
taken larger than the material loss of
$2\times10^{-4}$~\cite{Cagnoli1999Wires} to allow for additional
dissipation at joints and interfaces, the structural thermal noise
can dominate when
\begin{equation}
    k_s'
    <
    \frac{\phi_s k_0'^2}{k_b'\phi_b}
    \simeq
    \qty{5.8e9}{\newton\per\metre}.
\end{equation}
This illustrates the high support stiffness required to keep its
thermal-noise contribution below that of the diluted pendulum,
even though the suspension strongly attenuates the thermal force
generated by the support. In a multistage suspension, however, thermal noise generated in the upper supporting structure is additionally filtered by the mechanical transfer functions of the subsequent suspension stages. Its contribution is therefore generally negligible in the observation band of ground-based gravitational-wave detectors, which lies well above the relevant low-frequency suspension modes. However, this suppression does not apply at lower frequencies, particularly near or below the suspension resonances, or in instruments such as low-frequency torsion pendulums where the readout is intentionally sensitive in this regime.

In cases where structural thermal noise may be significant, a more detailed analysis of the supporting structure, including its geometry and attachment locations, may be warranted. Finite-element analysis can then be used to evaluate how structural deformation and resonant modes contribute to the readout thermal noise. Appendix~\ref{app:continuous} extends our source-based formulation to continuous systems and develops its modal representation, providing the framework for such an analysis.

\subsubsection{Rotational flexure on a lossy mount}

The same competition appears in rotational systems, such as the
configuration shown in Fig.~\ref{fig:rotational_flexure}, where a rigid
inertia is suspended by a low-loss flexure from a compliant supporting
structure.
\usetikzlibrary{
    arrows.meta,
    patterns,
    decorations.pathmorphing
}

\begin{figure}[ht]
\centering
\parbox[t]{0.47\linewidth}{\centering \bfseries\strut (a) Physical model}%
\hfill%
\parbox[t]{0.53\linewidth}{\centering \bfseries\strut (b) Lumped-element model}\par\vspace{2mm}
$\vcenter{\hbox{\resizebox{0.33\linewidth}{!}{%
\begin{tikzpicture}[
    scale=1,
    transform shape,
    >=Latex,
    line cap=round,
    line join=round,
    font=\small
]

\draw[line width=1pt] (-2.25,1.35)--(-2.25,2.45);

\foreach \y in {1.45,1.70,...,2.35}{
    \draw[line width=.55pt]
        (-2.25,\y)--++(-.20,.16);
}

\draw[line width=1pt] (2.25,1.35)--(2.25,2.45);

\foreach \y in {1.45,1.70,...,2.35}{
    \draw[line width=.55pt]
        (2.25,\y)--++(.20,.16);
}

\filldraw[
    fill=green!35,
    draw=green!50!black,
    line width=.9pt
]
    (-2.25,1.78) rectangle (2.25,2.00);

\node[
    green!45!black,
    align=center
] at (0,2.40)
{
    $k_s'(1+i\phi_s)$\\[-1pt]
    \textbf{structure}
};

\filldraw[
    fill=blue!35,
    draw=blue!70!black,
    line width=.7pt
]
    (-0.045,0.35) rectangle (0.045,1.78);

\node[
    blue!70!black,
    align=left,
    anchor=west
] at (0.28,1.05)
{
    $k_f'(1+i\phi_f)$\\[-1pt]
    \textbf{flexure}
};

\draw[
    black,
    line width=3pt
]
    (-0.90,0.35)--(0.90,0.35);

\filldraw[
    fill=black,
    draw=black
]
    (-0.90,0.35) circle (0.14);

\filldraw[
    fill=black,
    draw=black
]
    (0.90,0.35) circle (0.14);

\node[
    black,
    font=\small\bfseries
] at (0,0.1) {$J$};

\draw[
    densely dashed,
    line width=.8pt
]
    (-0.72,-0.3)
    arc[
        start angle=180,
        end angle=0,
        x radius=.72,
        y radius=.20
    ];

\draw[
    ->,
    line width=1pt
]
    (-0.72,-0.3)
    arc[
        start angle=180,
        end angle=320,
        x radius=.72,
        y radius=.20
    ];

\node at (-0.42,-0.78) {$\theta_{\scriptscriptstyle R}$};

\node[
    blue!70!black,
    anchor=west
] at (-0.22,-0.72)
{\textbf{readout}};

\end{tikzpicture}%
}}}$%
\hfill%
$\vcenter{\hbox{\resizebox{0.47\linewidth}{!}{%
\begin{tikzpicture}[
    scale=1,
    transform shape,
    >=Latex,
    line cap=round,
    line join=round,
    font=\small,
    spring/.style={
        decorate,
        decoration={
            zigzag,
            segment length=5pt,
            amplitude=3pt
        }
    }
]

\draw[line width=1pt] (0,-.75)--(0,.75);

\foreach \y in {-.65,-.4,...,.65}{
    \draw[line width=.55pt]
        (0,\y)--++(-.22,-.16);
}

\draw (0,0)--(.40,0);

\draw[
    green!50!black,
    spring,
    line width=.9pt
]
    (.40,0)--(1.85,0);

\draw (1.85,0)--(2.20,0);

\fill (2.20,0) circle (1.7pt);

\draw (2.20,0)--(2.55,0);

\draw[
    blue!70!black,
    spring,
    line width=.9pt
]
    (2.55,0)--(4.00,0);

\draw (4.00,0)--(4.40,0);

\filldraw[
    fill=white!16,
    draw=black,
    line width=1.2pt
]
    (4.40,-.42)
    rectangle
    (5.70,.42);

\node at (5.05,0) {$J$};

\node[
    green!45!black,
    align=center
] at (1.12,.82)
{$k_s'(1+i\phi_s)$};

\node[
    green!45!black,
    align=center
] at (1.12,-.92)
{\textbf{structure}};

\node[
    blue!70!black,
    align=center
] at (3.28,.82)
{$k_f'(1+i\phi_f)$};

\node[
    blue!70!black,
    align=center
] at (3.28,-.92)
{\textbf{flexure}};

\draw[
    ->,
    line width=1pt
]
    (5.95,0)--(6.95,0);

\node at (6.45,.28) {$\theta_{\scriptscriptstyle R}$};

\node[
    blue!70!black
] at (6.45,-.38)
{\textbf{readout}};

\end{tikzpicture}%
}}}$\par
\caption{
Rotational flexure mounted to a compliant supporting structure.
\textbf{(a)} Physical model consisting of a rigid inertia \(J\) suspended by a thin flexure from a supporting beam with finite rotational compliance.
\textbf{(b)} Equivalent lumped-element model, in which the structural stiffness \(k_s'(1+i\phi_s)\) and flexure stiffness \(k_f'(1+i\phi_f)\) act in series.
The readout is the angular displacement \(\theta_R\) of the rigid inertia.
}
\label{fig:rotational_flexure}
\end{figure}
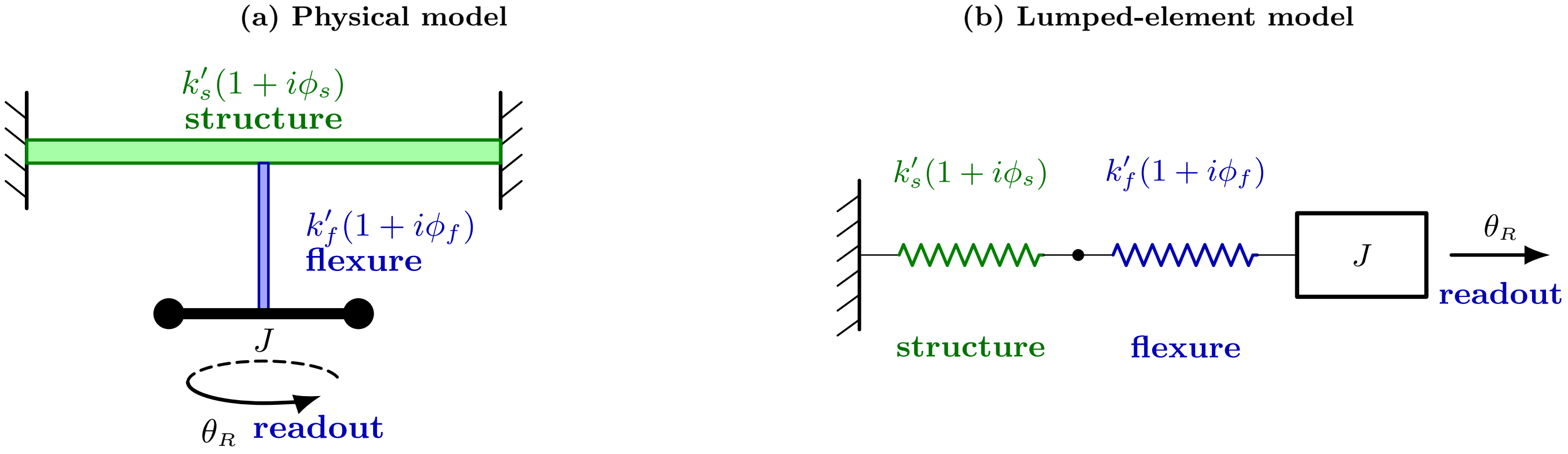

This model can be relevant to six-dimensional inertial sensors,
where additional compliance and dissipation may arise from the
structure supporting the suspension fiber of the Compact 6D
(C-6D) sensor~\cite{Smetana2025} or from a low-frequency bending mode of the arms
in the OmniSens system~\cite{Saffarieh2026}.
We consider a frequency range in which the flexure and structural
deformations remain quasistatic, retaining only the rotational
inertia of the reference mass. This inertia is connected to two
springs in series,
\begin{equation}
    k_f=k_f'(1+i\phi_f),
    \qquad
    k_s=k_s'(1+i\phi_s),
\end{equation}
where $k_f$ describes the flexure and $k_s$ represents the additional
structural stiffness.
We assume a low-loss flexure, such as a fiber with
$\phi_f\sim10^{-8}$, and a bolted aluminum arm or supporting
structure with $\phi_s\sim10^{-3}$.
The assumed loss angle of the support exceeds the structural loss of
$2.7\times10^{-4}$ measured in an aluminum flexure~\cite{Nguyen2015}
to account for additional dissipation at the joints.
At equal temperatures, \cref{eq:series ratio} shows that
the support dominates the thermal noise when
\begin{equation}
    \frac{k_s'}{k_f'}
    <
    \frac{\phi_s}{\phi_f}\sim10^5.
\end{equation}

\subsubsection{Rigidly mounted optics}

We now consider the unsuspended configuration shown in Fig.~\ref{fig:fixed_structure}, in which a mirror and its coating are mounted on a stiff supporting structure coupled to ground of finite compliance. We assume that all resonances of the support occur well above the frequency range of interest, such that the system can be described entirely by a series network of springs and no masses need to be considered. This configuration is relevant, for example, to unsuspended lunar interferometers such as LILA~\cite{Creighton2025LILA}.

\usetikzlibrary{
    arrows.meta,
    patterns,
    decorations.pathmorphing
}

\begin{figure}[tbp]
\centering
\parbox[t]{0.405\linewidth}{\centering \bfseries\strut (a) Physical model}%
\hfill%
\parbox[t]{0.535\linewidth}{\centering \bfseries\strut (b) Lumped-element model}\par\vspace{2mm}
$\vcenter{\hbox{\resizebox{0.405\linewidth}{!}{%
\begin{tikzpicture}[
    >=Latex,
    line cap=round,
    line join=round,
    font=\small
]

\draw[line width=1pt] (-3.2,-2.0)--(-3.2,1.55);
\draw[line width=1pt] (3.2,-2.0)--(3.2,1.55);
\draw[line width=1pt] (-3.2,-2.0)--(3.2,-2.0);

\foreach \y in {-1.9,-1.6,...,1.4}{
    \draw[line width=.5pt]
        (-3.2,\y)--++(-.20,.20);
    \draw[line width=.5pt]
        (3.2,\y)--++(.20,.20);
}

\foreach \x in {-3.0,-2.7,...,3.0}{
    \draw[line width=.5pt]
        (\x,-2.0)--++(-.20,-.20);
}

\fill[orange!12]
    (-3.15,-1.95) rectangle (3.15,-.25);

\foreach \x in {-3.0,-2.65,...,3.0}{
    \foreach \y in {-1.8,-1.45,...,-.4}{
        \fill[brown!45] (\x,\y) circle (.018);
    }
}

\draw[brown!60,line width=.6pt]
    (-3.15,-.25)--(3.15,-.25);

\foreach \x in {-.90,.90}{

    \filldraw[
        fill=gray!55,
        draw=black,
        line width=.6pt
    ]
        (\x-.045,-.25)
        rectangle
        (\x+.045,-.74);

    \foreach \y in {-.29,-.35,...,-.71}{
        \draw[line width=.4pt]
            (\x-.06,\y)--(\x+.06,\y+.025);
    }
}

\foreach \x in {-.90,.90}{

    \filldraw[
        fill=gray!40,
        draw=black,
        line width=.7pt
    ]
        (\x-.135,-.12)
        rectangle
        (\x+.135,.35);

    \filldraw[
        fill=gray!45,
        draw=black,
        line width=.7pt
    ]
        (\x-.30,-.25)
        rectangle
        (\x+.30,-.12);
}

\filldraw[
    fill=gray!35,
    draw=black,
    line width=.8pt
]
    (-1.55,.25)
    rectangle
    (1.55,.90);

\filldraw[
    fill=gray!45,
    draw=black,
    line width=.7pt
]
    (-.14,.90)
    rectangle
    (.14,1.18);

\filldraw[
    fill=blue!16,
    draw=black,
    line width=.8pt
]
    (-1.25,1.18)
    rectangle
    (1.18,1.62);

\filldraw[
    fill=orange!55,
    draw=black,
    line width=.8pt
]
    (1.18,1.18)
    rectangle
    (1.42,1.62);

\node[
    blue!70!black,
    align=center
] at (-.65,2.25)
{\textbf{mirror}\\[-1pt]$(k_m,\,T_m)$};

\draw[blue!70!black,->]
    (-.65,1.91)--(-.65,1.66);

\node[
    brown!70!black,
    align=center
] at (1.43,2.25)
{\textbf{coating}\\[-1pt]$(k_c,\,T_c)$};

\draw[brown!70!black,->]
    (1.28,1.91)--(1.28,1.66);

\draw[
    blue!70!black,
    ->,
    line width=1pt
]
    (1.42,1.40)--(2.25,1.40);

\node[
    black,
    anchor=west
] at (2.30,1.4)
{$x_{\scriptscriptstyle R}$};

\node[
    blue!70!black,
    anchor=west
] at (1.70,1.05)
{\textbf{readout}};

\node[black!70] at (0,0.58)
{\textbf{structure} $(k_s,\,T_s)$};

\node[
    pink!70!black,
    align=center
] at (-2.25,.13)
{\textbf{ground}\\[-1pt]$(k_G,\,T_G)$};

\draw[pink!70!black,->]
    (-2.15,-.20)--(-2.15,-.60);

\end{tikzpicture}%
}}}$%
\hfill%
$\vcenter{\hbox{\resizebox{0.535\linewidth}{!}{%
\begin{tikzpicture}[
    >=Latex,
    line cap=round,
    line join=round,
    font=\scriptsize,
    scale=1,
    spring/.style={
        decorate,
        decoration={
            zigzag,
            segment length=4.2pt,
            amplitude=2.6pt
        }
    }
]

\draw[line width=1pt]
    (0,-.7)--(0,.7);

\foreach \y in {-.6,-.35,...,.6}{
    \draw[line width=.55pt]
        (0,\y)--++(-.22,-.16);
}

\draw (0,0)--(.35,0);

\draw[
    pink!75!black,
    spring,
    line width=.9pt
]
    (.35,0)--(1.50,0);

\draw (1.50,0)--(1.82,0);
\fill (1.82,0) circle (1.6pt);

\draw (1.82,0)--(2.12,0);

\draw[
    black!75,
    spring,
    line width=.9pt
]
    (2.12,0)--(3.27,0);

\draw (3.27,0)--(3.59,0);
\fill (3.59,0) circle (1.6pt);

\draw (3.59,0)--(3.89,0);

\draw[
    blue!75!black,
    spring,
    line width=.9pt
]
    (3.89,0)--(5.04,0);

\draw (5.04,0)--(5.36,0);
\fill (5.36,0) circle (1.6pt);

\draw (5.36,0)--(5.66,0);

\draw[
    orange!75!black,
    spring,
    line width=.9pt
]
    (5.66,0)--(6.81,0);

\draw[
    blue!70!black,
    ->,
    line width=1pt
]
    (6.81,0)--(7.42,0)
    node[right,black] {$x_{\scriptscriptstyle R}$};

\node[
    pink!75!black,
    align=center
] at (.93,.60)
{$k_G'(1+i\phi_{\mathrm{G}})$\\[3pt]$T_G$};

\node[
    pink!75!black,
    align=center
] at (.93,-.60)
{\textbf{ground}\\[-1pt]\textbf{stiffness}};

\node[
    black!75,
    align=center
] at (2.70,.60)
{$k_s'(1+i\phi_s)$\\[3pt]$T_s$};

\node[
    black!75,
    align=center
] at (2.70,-.60)
{\textbf{supporting}\\[-1pt]\textbf{structure}};

\node[
    blue!75!black,
    align=center
] at (4.47,.60)
{$k_m'(1+i\phi_m)$\\[3pt]$T_m$};

\node[blue!75!black] at (4.47,-.60)
{\textbf{mirror}};

\node[
    orange!75!black,
    align=center
] at (6.24,.60)
{$k_c'(1+i\phi_c)$\\[3pt]$T_c$};

\node[orange!75!black] at (6.24,-.60)
{\textbf{coating}};

\node[blue!70!black] at (7.45,-.350)
{\textbf{readout}};

\end{tikzpicture}%
}}}$\par
\caption{
Fixed optical structure and equivalent series-spring model.
\textbf{(a)} Physical configuration of a mirror and coating mounted on a
supporting structure whose feet rest on compliant ground, with bolts
extending beneath the feet into the ground, and respective temperatures
\(T_m\), \(T_c\), \(T_s\), and \(T_G\).
\textbf{(b)} Equivalent lumped-element model, in which the ground,
supporting structure, mirror substrate, and coating are represented
by lossy stiffnesses connected in series at their respective temperatures.
The displacement \(x_R\) of the optical surface is taken as the readout.
}
\label{fig:fixed_structure}
\end{figure}
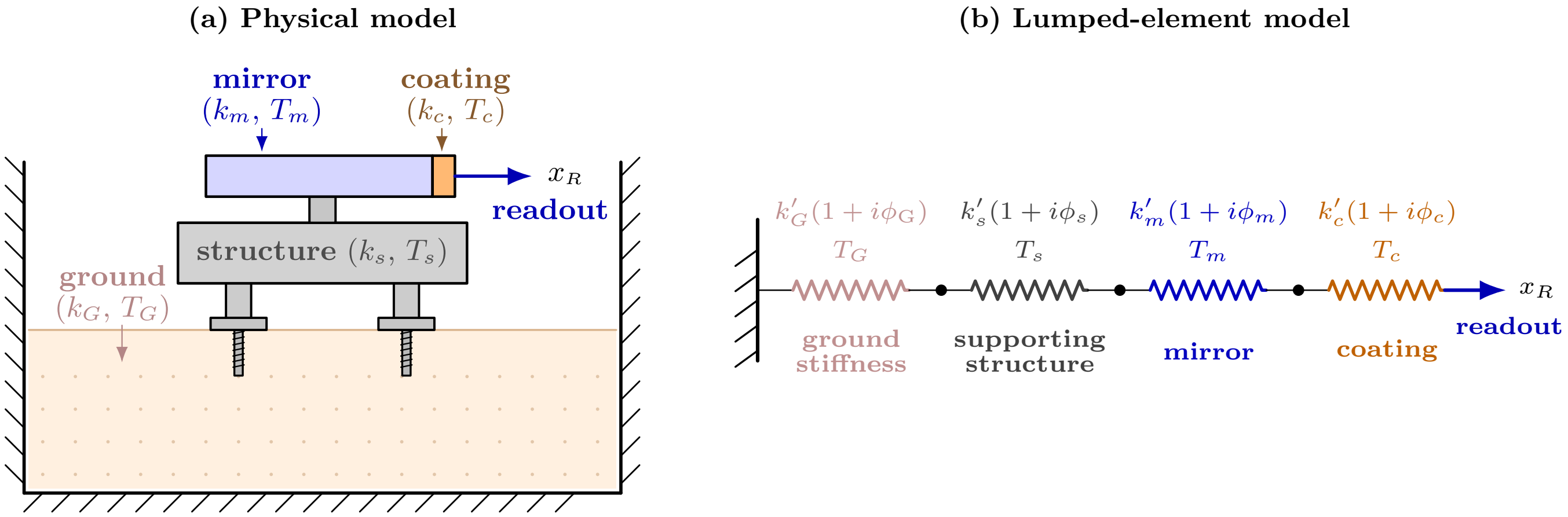

We model the system by four elastic elements in series, with frequency-independent elastic stiffnesses and structural loss angles. The ground and support springs describe
mechanical displacement, whereas the substrate and coating springs encode readout-weighted internal Brownian deformation, which appears as equivalent displacement sensing noise.

We assign each element a temperature \(T_j\), assuming local thermal equilibrium and approximately stationary temperatures. The optics may be maintained near $T_m \simeq T_c \simeq \qty{300}{\kelvin}$,
as envisaged for LILA~\cite{Creighton2025LILA}.
The lunar ground surface temperature $T_G$ is approximately
\qtyrange{250}{300}{\kelvin} near local noon and
\qtyrange{70}{80}{\kelvin} near local midnight~\cite{Williams2017}
for a representative site around $\qty{75}{\degree}$ latitude,
consistent with the latitude preference in iLILA's site-selection
study~\cite{Trippe2026}.

We assume that bolts anchor the support structure to the lunar ground
and that its contact area with the regolith has a characteristic
linear dimension $L_{\mathrm G}\sim\qty{10}{\centi\metre}$.
We model the regolith beneath the support as an elastic half-space
with shear modulus $G_{\mathrm G}$, loss angle $\phi_{\mathrm G}$,
and temperature $T_{\mathrm G}$. The static translational stiffness
associated with deformation of the regolith scales as~\cite{Gazetas1991}
\begin{equation}
    k'_{\mathrm G}\sim G_{\mathrm G}L_{\mathrm G},
\end{equation}
up to a dimensionless factor that depends on the contact geometry
and Poisson's ratio. Using the soil density at the Apollo 14
site~\cite{Mitchell1972} and near-surface shear-wave speeds inferred
from lunar seismic data~\cite{Amos2024}, we estimate
$G_{\mathrm G}=\rho v_s^2$ to be of order \qty{1e7}{\pascal}, giving
\begin{equation}
    k'_{\mathrm G}\sim\qty{1e6}{\newton\per\metre}.
\end{equation}

For an illustrative low-loss case, we use the intrinsic seismic
quality factor $Q_{\mathrm G}=6750$ adopted for the regolith by
Onodera et al.~\cite{Onodera2021LunarQ} and assume that the effective
loss angle beneath the support is constant over the frequency band
of interest, with
\begin{equation}
    \phi_{\mathrm G}\simeq Q_{\mathrm G}^{-1}
    \simeq 1.5\times10^{-4}.
\end{equation}

For the support, we consider an illustrative aluminum structure of
mass $m_s \sim \qty{10}{\kilogram}$ and height $L_s \sim \qty{1}{\metre}$.
We assume an optimistic configuration in which shear deformation dominates
the compliance over contributions from joints, rocking, or bending.
Using $\rho_s \simeq \qty{2700}{\kilogram\per\cubic\metre}$ gives an equivalent
cross-sectional area $A_s \sim m_s/(\rho_s L_s)
\simeq \qty{3.7e-3}{\square\metre}$.
With $G_s \simeq \qty{26}{\giga\pascal}$, the shear stiffness scales as
\begin{equation}
    k_s' \sim \frac{G_s A_s}{L_s} \sim \qty{1e8}{\newton\per\metre},
    \qquad \phi_s \sim 10^{-3}.
\end{equation}
Taking the aluminum structural loss to be of order
$2\times10^{-4}$~\cite{Nguyen2015}, we assume an effective
loss angle $\phi_s \sim 10^{-3}$ for the assembled structure
to allow for additional dissipation in bolted connections.

For the mirror substrate, we take the laser beam size to be \(w=\qty{0.5}{\milli\metre}\) as reported in
Ref.~\cite{Creighton2025LILA}, an elastic modulus
\(E\sim\qty{1e11}{\pascal}\), and a Poisson ratio
\(\nu\simeq0.17\), to arrive at
\begin{equation}
    k'_{\mathrm{m}}
    = \frac{\sqrt{\pi}Ew}{1-\nu^2}
    \sim\qty{1e8}{\newton\per\metre},
    \qquad
    \phi_{\mathrm{m}}\sim10^{-8}\text{--}10^{-7}
    \quad\text{\cite{Harry2002,Penn2006FusedSilica}}.
\end{equation}
The stiffness follows from the Gaussian half-space approximation,
which assumes that the substrate diameter and thickness are large compared with \(w\).
For high-quality fused-silica optics, the substrate thermal noise is known to be below the dielectric coating thermal noise in the gravitational-wave observation band~\cite{Harry2002,Penn2006FusedSilica}.

Finally, for a high-reflectivity silica--tantala coating of assumed
thickness \(d=\qty{5}{\micro\metre}\), with elastic moduli comparable to those
of the substrate, we take, to order of magnitude,
\begin{equation}
    k'_{\mathrm{c}}\sim\frac{Ew^2}{d}
    \sim\qty{5e9}{\newton\per\metre},
    \qquad
    \phi_{\mathrm{c}}\sim10^{-4}
    \quad\text{\cite{Harry2002}}.
\end{equation}
Clearly, treating the coating as a simple lossy spring glosses over many physical effects (e.g., field penetration into the coating and thermorefractive noise~\cite{Evans2008,Hong2013}), and yet it captures the dominant effect of ``coating Brownian noise,'' which is sufficient for this example. For comparison, rescaling our coating thermal-noise amplitude spectral density (ASD) to a beam radius of \(\qty{6.2}{\centi\metre}\) gives \(\qty{5.9e-21}{\metre\per\hertz\tothe{0.5}}\) at \(\qty{100}{\hertz}\), approximately \(\qty{18}{\percent}\) below the value of \(\qty{7.1e-21}{\metre\per\hertz\tothe{0.5}}\) obtained from the coating thermal-noise measurements of Gras \textit{et al.}~\cite{Gras2017} for the same beam radius, both evaluated at \(\qty{300}{\kelvin}\).
\begin{figure}[tbp]
    \centering
    \includegraphics[width=0.72\linewidth]{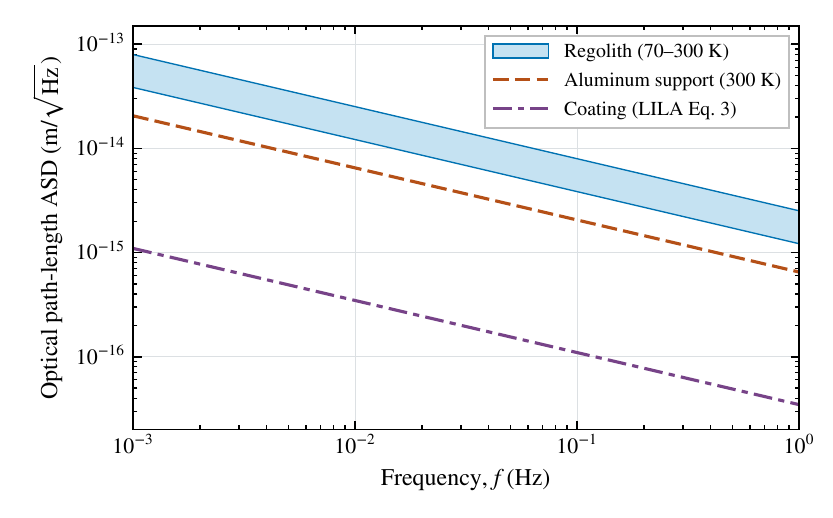}
    \caption{One-sided optical path-length amplitude spectral densities
    for rigidly mounted lunar optics. The coating curve uses
    Eq.~(3) of Ref.~\cite{Creighton2025LILA}.
    The support and regolith curves use the series-spring model
    with a common ASD multiplier of $2\sqrt{4}$ for four independent,
    equally coupled displacements in a round-trip optical path.
    The support temperature is \qty{300}{\kelvin}. The regolith
    band shows \qtyrange{70}{300}{\kelvin}.}
    \label{fig:lila_thermal_noise_comparison}
\end{figure}
In the low-frequency limit, the series-spring result derived previously reduces to
\begin{equation}
    S_{x_R}(\omega)
    \simeq
    \frac{4k_B}{\omega}
    \sum_{j\in\{\mathrm{G},\mathrm{s},\mathrm{m},\mathrm{c}\}}
    T_j\frac{\phi_j}{k_j'}.
\end{equation}
Taking the coating thermal noise as a reference, the corresponding figures of merit are
\begin{equation}
\frac{\phi_{\mathrm{c}}}{k'_{\mathrm{c}}}
\sim 2\times10^{-14},
\qquad
\frac{T_m}{T_c}\frac{\phi_{\mathrm{m}}}{k'_{\mathrm{m}}}
\sim 10^{-16}\text{--}10^{-15},
\qquad
\frac{T_s}{T_c}\frac{\phi_{\mathrm{s}}}{k'_{\mathrm{s}}}
\sim \frac{T_s}{T_c}10^{-11},
\qquad
\frac{T_G}{T_c}\frac{\phi_{\mathrm{G}}}{k'_{\mathrm{G}}}
\sim \frac{T_G}{T_c}10^{-10}.
\end{equation}

Even though the lunar ground is itself low loss and possibly colder
than the optics, its larger compliance can make its thermal-noise
contribution much larger than those of the mirror substrate and
coating. A similar conclusion can be drawn for the supporting
structure. 
To compare with the coating benchmark in Eq.~(3) of
Ref.~\cite{Creighton2025LILA}, we adopt the same output mode cleaner
normalization and multiply the modeled support and
regolith displacement ASDs by $2\sqrt{4}$, accounting for the round trip
and four independent contributions associated with the four optics.
Figure~\ref{fig:lila_thermal_noise_comparison} shows that, for the
illustrative parameters considered here, the regolith ASD could exceed
the coating benchmark by a factor of $\sim50$ at representative
temperatures, and the support ASD by a factor of $\sim20$.

\subsection{Feedback control as a nonthermal force and electrical analogy}

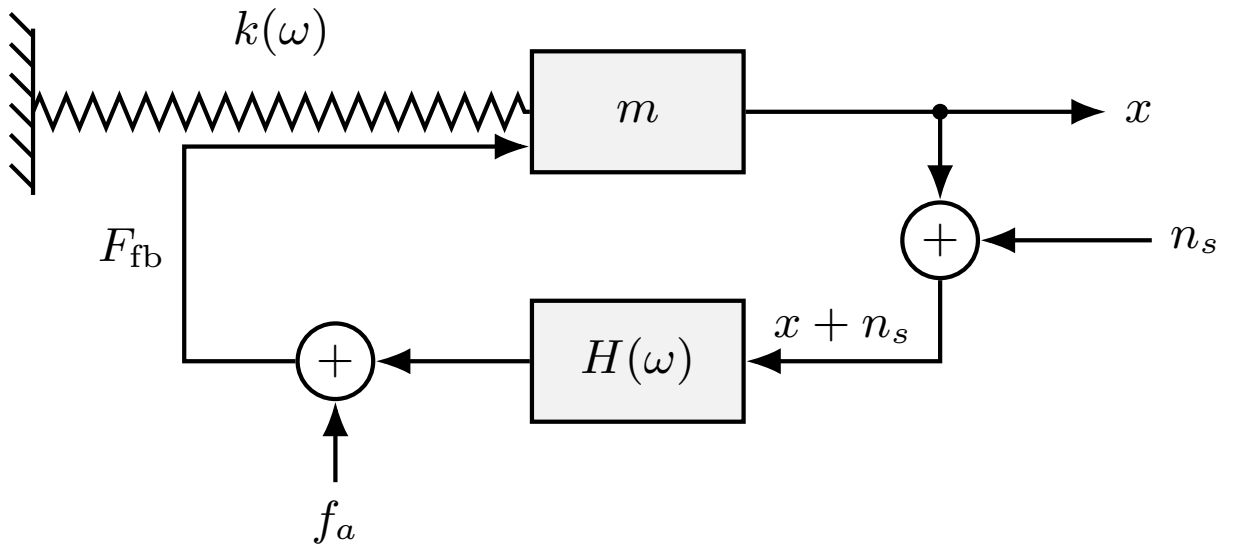
\begin{figure}[htbp]
\centering
\begin{tikzpicture}[>=Latex,line width=.8pt,
 spring/.style={decorate,decoration={zigzag,segment length=5pt,amplitude=3pt}},
 block/.style={draw,minimum width=1.4cm,minimum height=.8cm,fill=gray!10},
 sum/.style={draw,circle,inner sep=0pt,minimum size=5mm}]
\draw (0,-.55)--(0,.55);
\foreach \y in {-.5,-.3,...,.5} \draw (0,\y)--++(-.15,.15);
\node[block] (mass) at (4,0) {$m$};
\draw[spring] (0,0)--(mass.west) node[midway,above=7pt] {$k(\omega)$};
\draw[->] (mass.east)--(7.1,0) node[right] {$x$};
\fill (6,0) circle (1.5pt);
\node[sum] (sense) at (6,-.85) {$+$};
\draw[->] (6,0)--(sense.north);
\draw[->] (7.4,-.85) node[right] {$n_s$}--(sense.east);
\node[block] (controller) at (4,-1.65) {$H(\omega)$};
\draw[->] (sense.south)--(6,-1.65)--(controller.east) node[midway,above] {$x+n_s$};
\node[sum] (act) at (2,-1.65) {$+$};
\draw[->] (controller.west)--(act.east);
\draw[->] (2,-2.45) node[below] {$f_a$}--(act.south);
\draw[->] (act.west)--(1,-1.65)--(1,-.23)--([yshift=-2.3mm]mass.west);
\node[left] at (1,-.9) {$F_{\rm fb}$};
\end{tikzpicture}
\caption{A suspended mass with noisy feedback. Displacement-referred sensing noise $n_s$ enters before the controller; additive actuation force noise $f_a$ enters after it.}
\label{fig:thermal_feedback}
\end{figure}

Furthermore, this approach naturally accommodates nonthermal noise sources, including sensing and actuation noise in feedback systems and quantum radiation-pressure noise in optomechanical systems~\cite{Caves1980a,instrO3paper}. Such interactions can modify the mechanical response and provide damping without the thermal-force noise associated with equivalent passive dissipation. Their additional noise sources must therefore be treated separately.
For example, consider a mass $m$ with displacement $x$, suspended
by a lossy spring $k(\omega)$ at temperature $T$, with linear
time-invariant feedback as shown in Fig.~\ref{fig:thermal_feedback}. If $n_s$ is displacement-referred sensing noise and $f_a$ is
additive actuation force noise, the force applied to the mass by the feedback
actuator is
\begin{equation}
 F_{\rm fb}(\omega)=H(\omega)\bigl[x(\omega)+n_s(\omega)\bigr]+f_a(\omega).
\end{equation}
We choose $H\om$ so that the closed-loop dynamics are asymptotically
stable. The closed-loop dynamic compliance is

\begin{equation}
 C_{\rm fb}(\omega)=\frac{1}{k(\omega)-m\omega^2-H(\omega)}.
\end{equation}
If the spring's thermal force, sensing noise, and actuation noise are mutually uncorrelated, the physical displacement PSD is
\begin{equation}
 S_x^{\rm fb}(\omega)=|C_{\rm fb}(\omega)|^2
 \left[\frac{4k_BT}{\omega}\imag{k(\omega)}
 +|H(\omega)|^2S_{n_s}(\omega)+S_{f_a}(\omega)\right].
\end{equation}
Feedback changes the response while leaving the spring's thermal-force noise unchanged. This assumes
that externally powered feedback does not change the spring’s intrinsic dissipation or temperature. Here, the quantity of interest is the physical displacement $x$,
not the sensor output $x+n_s$. We therefore intentionally use $x$, $S_x$, and
$C_{\rm fb}$ to avoid interpreting the readout label $R$ as referring
to the noisy sensor output.

Applying Saulson's relation~\cite{Saulson1990jc},
given in \cref{eq:saulson_in_our_notations}, directly to the
closed-loop compliance would assign the controller a fictitious
thermal-force PSD $-(4k_BT/\omega)\imag{H(\omega)}$.

The treatment of thermal-force noise developed here has a direct analogue in electrical circuit theory. In the impedance analogy, mechanical force corresponds to voltage and velocity to current, so the thermal-force noise associated with a lossy elastic element $\Sfj$ corresponds to the Johnson--Nyquist voltage noise of a resistor.
In the mobility analogy, force corresponds to current and velocity to voltage, so the same thermal-force noise corresponds to the Norton current-noise representation of the resistor
~\cite{Hunt1954,Nyquist1928}. This analogy is useful beyond simple interpretation. Electrical noise analysis routinely represents dissipative elements by local noise sources and propagates their contributions through a linear network to a specified readout. The resulting framework supports arbitrary network topologies, multiple noise sources, frequency-dependent impedances, and both voltage- and current-like readouts. A substantial body of circuit-analysis techniques and numerical tools has therefore been developed for precisely this type of problem. The formulation used here may be viewed as the mechanical counterpart of that established approach: thermal fluctuations are assigned to the individual dissipative elements and propagated through the mechanical system to the observable of interest.

\section{Conclusion}
\label{sec:conclusion}

We have presented a systematic thermal-force source description for
calculating and interpreting thermal noise in lumped-element mechanical
systems. Expressing the readout noise as a sum of individual contributions
makes explicit both the noise generated by each dissipative element and
its transmission through the mechanical system.

We used this description to clarify why parallel and series dissipation
dilution reduce displacement noise in different frequency regimes.
Through several examples, we showed that thermal noise generated by
supporting structures can dominate the readout noise even when those
structures have negligible influence on the system dynamics. For a
feedback-controlled suspension, the same source-based approach
separates the modification of the mechanical response by the controller
from the noise introduced by sensing and actuation.

These examples demonstrate how source-by-source comparisons can
identify potential design limitations and guide optimization before
undertaking detailed structural modeling or finite-element analysis.
Even when a reduced model does not support a definitive design
decision, it can identify which components or mechanisms warrant
further investigation. This makes the method particularly useful
during the early design stages of precision mechanical instruments.

\appendix

\section{Formalism and derivation of the transfer functions}
\label{sec:notation}

This appendix lays out the notation used throughout this work for describing
mechanical systems composed of lumped-element stiffnesses and pointlike masses.
We represent a mechanical system as a graph with nodes
and edges (connections between nodes).
Masses are associated with nodes, while springs, flexures, dampers, or other generalized stiffness elements are associated with edges.
Massless nodes may be included
when needed to describe elements in series. %
We lay out our choice of variable names in \cref{tab:variable names}.

\begin{table}[ht]
\centering
\begin{tabular}{lll}
\hline
\textbf{Name} & \textbf{Symbol} & \textbf{Comment} \\
\hline
\multicolumn{3}{l}{\textbf{Dynamic vectors}} \\
Force & $\vc{f}$ & Generalized effort variable \\
Position & $\vc{x}$ & Generalized coordinate \\
Shape & $\vpsi$ &  Connects generalized variables \\
Stretch & $\vc{y}$ & Elongation of elastic elements \\
Connection vector & $\vc{e}$ &  Connects the elastic elements \\
\hline
\multicolumn{3}{l}{\textbf{Element properties}} \\
Mass & $m$ &  Inertial element property \\
Generalized stiffness &$k = k' + i k''$ & Elastic element property \\
Restoring stiffness &$k' = \real{k}$ & Conservative response \\
Damping &$k'' = \imag{k} = \omega b$ & Dissipative response \\
Loss angle &$\phi = k'' / k'$ &  Fractional dissipation \\
\hline
\multicolumn{3}{l}{\textbf{System matrices}} \\
Mass &$\mM$ & Mass matrix in nodal basis \\
Generalized stiffness &$\mK = \mE \, \mKe \, \mE^T$ & Stiffness between nodes \\
Restoring stiffness &$\mK'$ & $\real{\mK}$ \\
Damping &$\mK''$ & $\imag{\mK}$ \\
Dynamic stiffness &$\vc{f} =\mD\vc{x} $ & $\mD = \mK - \omega^2 \mM$ \\
Dynamic compliance &$\vc{x}=\mC \vc{f}$ & $\mC = \mD^{-1}$ \\
\hline
\multicolumn{3}{l}{\textbf{System scalars}} \\
Stored energy sum & $U$ & Sum of the individual elements' peak elastic strain energies\\
Dissipated power & $W_{\rm diss}$ & Dissipation in stiffness elements\\
Displacement noise & $\Sx$ & Power spectral density (PSD) in $\unit{\square\metre\per\hertz}$ \\
Force noise & $\Sf$ & PSD in $\unit{\square\newton\per\hertz}$ \\
\hline
\end{tabular}
\caption{Definitions of variables.}
\label{tab:variable names}
\end{table}

Unless stated otherwise, we use lowercase letters for scalar quantities, bold lowercase letters for vectors, and capital letters for matrices.
The superscript $T$ is used for the transpose operation, the $*$ for conjugation, and $\dagger$ for the conjugate transpose (i.e., $\tc{\vpsi} = (\vpsi^*)^T$).

The elastic elements of our system will be labeled using the letter $j$ in the subscript or superscript (i.e., $k_j$ is the scalar stiffness of element $j$).
This label does not, however, specify the location of the named property in the associated system matrix, since this depends on how elements are \emph{connected} (see below).
For vectors and matrices, the subscript $e$ will be used to indicate that the vector is a list of element properties, or the corresponding diagonal \emph{element} matrix.
For instance,
$\vc{k}_e = [k_1, k_2, \dots]^T$ is the vector of element stiffnesses, and $\mKe = \diag{\vc{k}_e}$ is the corresponding diagonal matrix.

System matrices are generally \emph{not} in the element basis, but rather in the basis described by the nodes of the system graph.
When it is necessary to specify the value of a matrix or vector component, subscripts
$a$ and $b$ will be used for rows and columns, as in $\mK_{ab}$ for the generalized stiffness between nodes $a$ and $b$.
When \emph{both element labels and matrix indices} are required, the element label $j$ is moved to the superscript, as in
\begin{equation}
    \mK_{ab}
    = \sum_j \mK_{ab}^j
    = \sum_j \ve_a^j \, k_j \, \ve_b^j
    = \left(\mE   \, \mKe \,\mE ^T\right)_{ab},
\end{equation}
which indicates, for instance,
that the generalized stiffness matrix component $\mK_{ab}$ is the sum of the stiffnesses of all elements connected between nodes $a$ and $b$ in the system graph.
See \cref{sec:graph} below for more details about the connection graph implementation in matrix form.
The superscript notation for element labels may also be used for scalar quantities where subscripts might cause confusion (e.g., $\sbsp{W}{\rm diss}{j}$ for the average power dissipated by element $j$).

The analysis done in this paper is based on the standard frequency-domain treatment of linear systems found in any mechanics textbook \cite{textbook:Inman1994,textbook:Meirovitch1997,textbook:Cook2007FEA}.
We define a generalized coordinate $x_A$ as a linear combination
of the nodal displacements,
\begin{equation}
    x_A(\omega)=\vpsi_A^T\vx(\omega),
    \qquad
    \vpsi_A=
    \begin{bmatrix}
        \psi_{A,1} & \psi_{A,2} & \cdots & \psi_{A,N}
    \end{bmatrix}^T,
\end{equation}
where $\vpsi_A$ is the shape vector defining coordinate $A$.
The scalar $x_A(\omega)$ is the complex amplitude of harmonic motion,
such that
\begin{equation}
    x_A(t)=\real{x_A(\omega)e^{i\omega t}}.
\end{equation}
The same notation is used for all vector quantities.  For instance, the vector associated with
 generalized force ``A'' is  $\vc{f}_A\om = f_A(\omega) \vpsi_A$.
Thus, the label ``R'' is used for the readout variable.  The generalized coordinate defined by the readout $x_R$ has shape vector $\vpsi_R$, and the generalized force $\vf_R$ drives its generalized conjugate momentum.

\subsection{Mechanical connection graph}
\label{sec:graph}
In a system with $N$ nodes and $N_e$ stiffness elements, let
$\vx$ be the $N$-component vector of nodal displacements and $\vc{y}$ the $N_e$-component vector
of element elongations. Each element $j$ is associated with an
$N$-component connection vector $\ve_j$ such that its elongation is written
\begin{equation}
    y_j = \ve_j^T \vx.
\end{equation}
For example, if element \(j\) connects node \(a\) to node \(b\), then
\begin{equation}
    y_j = x_b - x_a,
\end{equation}
which implies \(\ve_a^j = -1\) and 
\(\ve_b^j = +1\) (with all other components of $\ve_j$ equal to zero).
Collecting all of the element connection vectors gives the connection matrix
\begin{equation}
    \vc{y} = \mE^T \vc{x},
\end{equation}
where $\mE$ is an $N \times N_e$ matrix whose columns are the vectors
$\vc{e}_j$ (i.e., $\mE = [\vc{e}_1,\ldots,\vc{e}_{N_e}]$).
With the connections between stiffness elements defined, we can now build
the system operators listed in \cref{tab:variable names}.
We start with a diagonal element-by-element stiffness matrix,
\begin{equation}
    \mKe(\omega)
    =
    \operatorname{diag}\!\left[
        k_1(\omega), k_2(\omega), \dots, k_{N_e}(\omega)
    \right]
    =
    \mKe'(\omega) + i\,\mKe''(\omega),
\end{equation}
where
\begin{equation}
    \mKe'(\omega)
    =
    \operatorname{Re}\!\left[\mKe(\omega)\right],
    \qquad
    \mKe''(\omega)
    =
    \operatorname{Im}\!\left[\mKe(\omega)\right].
\end{equation}

The nodal stiffness matrix is then
\begin{equation}
   \mK \om  =\mE \, \mKe(\omega) \,\mE ^T. 
    \label{eq:K_from_E}
\end{equation}
The elements can reach peak elastic strain energy at different times within a cycle. We denote the sum of their individual peak energies by $U$.

\begin{equation}
    U\om \equiv \sum_j \Umax^j\om
    =
    \frac12 \vy^\dagger\om \, \mKe'\om \, \vy\om
    =
    \frac12 \vx^\dagger\om \, \mK'\om \, \vx\om.
\end{equation}
The mass matrix $\mM$ is similarly constructed, with mass elements contributing 
to diagonal components. Note that $\mM$ could be frequency dependent. For a vector of applied nodal forces $\vf(\omega)$, the
frequency-domain equations of motion are
\begin{equation}
    \vf\om = \mD\om \, \vx\om
    \quad\text{with solution}\quad
    \vx\om = \mC\om \, \vf\om,
\end{equation}
where the dynamic stiffness and compliance matrices are
\begin{equation}
    \mD\om = \mK\om - \omega^2 \mM
    \quad\text{and}\quad
    \mC\om = \mD^{-1}(\omega).
\end{equation}
The role of the connection matrix is therefore twofold.
First, $\mE^T$ maps \emph{nodal coordinates to element elongations},
\(\vc{y} =\mE ^T  \vc{x}\),
thereby encoding the connections between nodes.
Second, $\mE$ maps \emph{forces across elements to forces on nodes},
\(\vc{f} =\mE \, \vc{f}_e\).
This second use encodes the fact that a spring force is applied with opposite signs to the two nodes at its ends.

\subsection{General calculation of transfer factors}
\label{sec:transfer factors}

In this appendix, we give a systematic construction of the transfer factors
\(G_j(\omega)\) appearing in \cref{eq:Sx B}, for the case where the readout variable
is any generalized coordinate of the system. Using the connection matrix defined above, the nodal displacement response
to the vector of element thermal forces $\vf_e(\omega)$ is
\begin{equation}
    \vx(\omega) = \mC(\omega)\mE\vf_e(\omega).
\end{equation}
The transfer function from the element-generated force
$f_j = (\vf_e)_j$ to the generalized coordinate $x_A$ defined by
$\vpsi_A$ is the transfer compliance
\begin{equation}
    \frac{x_A\om}{f_j\om}
    =
    \CAj\om
    \qquad\text{where}\qquad
    \CAj\om \equiv \CAjdef.
    \label{eq:coord_element_compliance}
\end{equation}
On the other hand, the transfer function from a generalized force
$f_A$, applied through the nodal force vector $\vf_A = \vpsi_A f_A$,
to $x_A$ is the self-compliance
\begin{equation}
    \frac{x_A(\omega)}{f_A(\omega)}
    = C_A(\omega)
    \qquad\text{where}\qquad
    C_A(\omega) \equiv \vpsi_A^T\mC(\omega)\vpsi_A.
\end{equation}
Requiring both forces to produce the same displacement gives the
force-transfer factor
\begin{equation}
    \sbsp{G}{A}{j}\om
    \equiv
    \frac{f_A\om}{f_j\om}
    =
    \frac{\CAj\om}{C_A\om}.
    \label{eq:Gaj_def}
\end{equation}
To reduce the notation overhead for the primary result, we define
\begin{equation}
    G_j\om \equiv G_R^j\om
    \quad\text{and}\quad
    C_j\om \equiv C_R^j\om,
    \label{eq:Gj_def}
\end{equation}
for the system readout variable $x_R$. This separates each element's force transfer from the readout compliance,
making the relative contributions of different force paths to the thermal
noise easier to interpret.

\subsection{Thermal-noise transfer}
\label{sec:noise transfer}

As shown in \cref{sec:derivation}, the force-noise power spectrum associated with
element \(j\) is
\begin{equation}
    \Sfj\om
    =
    \frac{4k_B T_j}{\omega}
    \operatorname{Im}\!\left[
    k_j(\omega)  \right],
\end{equation}
and, since each of these noise sources is independent, the displacement noise of coordinate \(x_A\) is
\begin{equation}
    S_{x_A}(\omega)
    =
    \sum_{j=1}^{N_e}
    \left|
    \CAj\om
    \right|^2
    \Sfj\om.
    \label{eq:Sxa_direct}
\end{equation}
Using \cref{eq:Gaj_def}, this can be written in the form of \cref{eq:Sx B},
\begin{equation}
    \SxA(\omega)
    =
    |C_{A}(\omega)|^2
    \sum_{j=1}^{N_e}
    |\sbsp{G}{A}{j}\om|^2
    \Sfj\om.
    \label{eq:Sxa_CG}
\end{equation}
The thermal-noise spectra for all nodal coordinates can similarly be expressed by defining
a vector of element force-noise spectra
\begin{equation}
    \Svf (\omega)
    =
    \begin{bmatrix}
    S_f^1(\omega)\\
    S_f^2(\omega)\\
    \vdots\\
    S_f^{N_e}(\omega)
    \end{bmatrix},
\end{equation}
and a vector of coordinate displacement-noise spectra
\begin{equation}
    \Svx(\omega)
    =
    \begin{bmatrix}
    S_{x_1}(\omega)\\
    S_{x_2}(\omega)\\
    \vdots\\
    S_{x_N}(\omega)
    \end{bmatrix}.
\end{equation}
Then \cref{eq:Sxa_direct} can be generalized to a multi-coordinate version of \cref{eq:Sx B} as
\begin{equation}
    \Svx(\omega)
    =
    \dsqr{\mC \om\mE}
    \Svf(\omega),    
    \label{eq:sq_direct}
\end{equation}
where \(\dsqr{\cdot}\) denotes the element-wise squared modulus.

\section{Derivation of the thermal-force source formulation}
\label{sec:derivation}

The Levin proof below assumes a common temperature $T$ and a passive, stable, reciprocal linear
system with real readout weighting. All spectra are one-sided PSDs per hertz, evaluated at
angular frequency $\omega$.
While \cref{eq:Sx A} is supported by the Langevin equation,
which serves as the starting point for Saulson's analysis in
\cite{Saulson1990jc}, we will show in this section that our main result
can also be derived from the fluctuation--dissipation theorem by
following Levin's derivation in \cite{levin1998}.

We use the nodal compliance and connection vectors introduced with \cref{eq:main_nodal_response,eq:main_source_compliance} and Appendix~\ref{sec:notation}.

We start with a generalized ``virtual force'' $f_R\om$
applied to the system at the generalized position coordinate $x_R$,
which serves as the ``readout variable'' for our system.
To be explicit, for a readout with
a shape vector (or ``form factor'' in Levin's work)
$\vpsi_R$, the readout displacement and virtual force vector
are
\begin{equation}
    x_R\om =  \vpsi_R^T\vx\om
    \quad \text{and} \quad
    \vf_R(\omega) = f_R\om \, \vpsi_R.
\end{equation}
If $\Wdiss$ denotes the time-averaged power dissipated by the system
when driven by $f_R$,
then Levin's result can be written in our notation as
\begin{equation}
\label{eq:levin_omega}
    \SxR\om
    =
    \frac{8 \kBT}{\omega^2}
    \frac{\Wdiss\om}{   
        \;\asqr{f_R\om}  }.
\end{equation}
We note that our formalism can be generalized to continuous systems
by exchanging matrix products for overlap integrals with
$\vpsi_n \rightarrow \vpsi_n(\vec{r})$,
as in Eq.~(3) of Levin~\cite{levin1998},
but we will avoid the associated notational complexity.

Figure~\ref{fig:thermal-reciprocity} illustrates these definitions
for a system of three masses and springs and shows the two equivalent driving configurations for calculating the thermal noise produced by the system.

\usetikzlibrary{arrows.meta,decorations.pathmorphing}

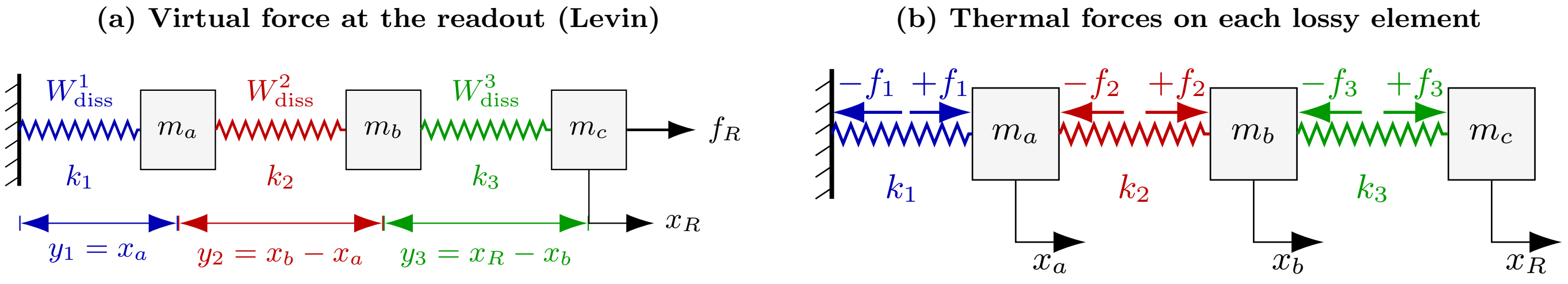
\begin{figure}[ht]
\centering
\def\drawForcePanel#1{\def\panel{#1}%
\begin{tikzpicture}[
    >={Latex[length=3mm,width=2mm]},
    font=\small,
    mass/.style={
        draw,fill=gray!8,
        minimum width=.8cm,
        minimum height=.85cm
    },
    spring/.style={
        thick,decorate,
        decoration={zigzag,segment length=4pt,amplitude=2.5pt}
    },
    elemzero/.style={blue!70!black},
    elemone/.style={red!75!black},
    elemtwo/.style={green!60!black}
]

\begin{scope}

\draw[very thick] (0,-.6)--(0,.6);
\foreach \h in {-.5,-.3,-.1,.1,.3,.5}{
    \draw (-.15,\h-.1)--(0,\h);
}

\node[mass] (a) at (1.7,0) {$m_a$};
\node[mass] (b) at (3.9,0) {$m_b$};
\node[mass] (c) at (6.1,0) {$m_c$};

\draw[spring,elemzero] (0,0)--(a.west);
\draw[spring,elemone] (a.east)--(b.west);
\draw[spring,elemtwo] (b.east)--(c.west);

\node[elemzero] at (.65,-.5) {$k_1$};
\node[elemone] at (2.8,-.5) {$k_2$};
\node[elemtwo] at (5,-.5) {$k_3$};

\ifnum\panel=0

\node[elemzero] at (.65,.42) {$W_{\mathrm{diss}}^1$};
\node[elemone] at (2.8,.42) {$W_{\mathrm{diss}}^2$};
\node[elemtwo] at (5,.42) {$W_{\mathrm{diss}}^3$};

\draw[->,thick] (c.east)--(7.25,0)
    node[right] {$f_R$};

\draw[|<->|,elemzero] (0,-1)--(1.7,-1)
    node[midway,below=3pt] {$y_1=x_a$};
\draw[|<->|,elemone] (1.7,-1)--(3.9,-1)
    node[midway,below=3pt] {$y_2=x_b-x_a$};
\draw[|<->|,elemtwo] (3.9,-1)--(6.1,-1)
    node[midway,below=3pt] {$y_3=x_R-x_b$};

\draw[->] (c.south)--(6.1,-1)--(6.8,-1)
    node[right] {$x_R$};

\else

\draw[->,thick,elemzero] (.65,.2)--(0,.2)
    node[midway,above] {$-f_1$};
\draw[->,thick,elemzero] (.72,.2)--(a.west |- 0,.2)
    node[midway,above] {$+f_1$};

\draw[->,thick,elemone] (2.7,.2)--(a.east |- 0,.2)
    node[midway,above] {$-f_2$};
\draw[->,thick,elemone] (2.9,.2)--(b.west |- 0,.2)
    node[midway,above] {$+f_2$};

\draw[->,thick,elemtwo] (4.9,.2)--(b.east |- 0,.2)
    node[midway,above] {$-f_3$};
\draw[->,thick,elemtwo] (5.1,.2)--(c.west |- 0,.2)
    node[midway,above] {$+f_3$};

\foreach \name/\pos/\label in {a/1.7/x_a,b/3.9/x_b,c/6.1/x_R}{
    \draw[->] (\name.south)--(\pos,-1)--(\pos+.65,-1)
        node[midway,below] {$\label$};
}

\fi
\end{scope}
\end{tikzpicture}%
}
\parbox[t]{.48\linewidth}{\centering \bfseries\strut (a) Virtual force at the readout (Levin)}%
\hfill%
\parbox[t]{.48\linewidth}{\centering \bfseries\strut (b) Thermal forces on each lossy element}\par\vspace{2mm}
$\vcenter{\hbox{\resizebox{.48\linewidth}{!}{%
\drawForcePanel{0}%
}}}$%
\hfill%
$\vcenter{\hbox{\resizebox{.48\linewidth}{!}{%
\drawForcePanel{1}%
}}}$\par
\caption{Two equivalent driving configurations for calculating thermal noise in the same three-mass system.
In this example, $\vx=(x_a,x_b,x_R)^T$, and the dynamic compliance
matrix is
$\mC(\omega)=
\left(
\begin{smallmatrix}
k_1+k_2-m_a\omega^2 & -k_2 & 0 \\
-k_2 & k_2+k_3-m_b\omega^2 & -k_3 \\
0 & -k_3 & k_3-m_c\omega^2
\end{smallmatrix}
\right)^{-1}$,
where $k_j=k_j(\omega)$ are the complex element stiffnesses. The corresponding connection vectors are $\ve_1^T=(1,0,0)$, $\ve_2^T=(-1,1,0)$, and $\ve_3^T=(0,-1,1)$.
(a) A virtual force $f_R$ applied at the readout produces element
elongations $y_j$ and dissipated powers $\Wdiss^j$.
(b) Each element generates a thermal force $f_j$ that acts with equal and
opposite signs at its ends and produces a displacement at the readout
$x_R$ through the compliance $C_j(\omega)$.
}

\label{fig:thermal-reciprocity}
\end{figure}

For a system composed of several elements, the total dissipated power
is the sum of the power dissipated in each element:
\begin{equation}
    \Wdiss\om
    =
    \sum_j \Wdiss^j\om.
\end{equation}
If element $j$ is represented by a complex generalized stiffness
\begin{equation}
    k_j\om = k'_j\om + i k''_j\om
    = k'_j\om (1 + i \phi_j\om),
\end{equation}
and $y_j$ is the elongation or ``stretch'' of element $j$
under the applied virtual force $f_R$,  then the power dissipated
in that element, averaged over one cycle, is
\begin{equation}
    \Wdiss^j\om
    =
    \frac12 \omega \, k''_j\om \asqr{y_j\om}.
    \label{eq:Wdiss_j}
\end{equation}
Substituting \cref{eq:Wdiss_j} into \cref{eq:levin_omega},
and using the definition of $\Sfj$ in \cref{eq:SFj}, gives
\begin{equation}
    \SxR\om
    =
    \sum_j
    \asqr{ \frac{y_j\om}{f_R\om} }
    \Sfj\om.
    \label{eq:levin_sum}
\end{equation}
Thus Levin's method is equivalent
to
propagating each
element's
thermal-force noise to the readout
with the compliance \emph{from} the virtual force applied at the
readout \emph{to} the displacement across element $j$.
Mechanical reciprocity ($\mC = \mC^T$) gives
\begin{equation}
\frac{y_j\om}{f_R\om}
= \ve_j^T \mC\om \vpsi_R
= \vpsi_R^T \mC\om \ve_j
= \frac{x_R\om}{f_j\om}
= \CRj\om
\equiv \Cj\om,
\end{equation}
where each ratio is evaluated with only its corresponding driving
force applied,
such that \cref{eq:levin_sum} can also be written as
\begin{equation}
    \SxR\om
    =
    \sum_j
    \asqr{ \frac{x_R\om}{f_j\om} }
    \Sfj\om,
    \label{eq:thermal_force_sum}
\end{equation}
which is our \cref{eq:Sx A},
and matches the description of $\Cj$ as
``compliance that connects a force across element $j$ to the
readout variable.''

For lumped elements at different steady temperatures in local thermal
equilibrium, the same derivation applies with the following substitution
in \cref{eq:levin_omega}, obtained from the discrete form of Eq.~(1) of
Komori et al.~\cite{Komori2018}:
\[
    T \Wdiss\om \;\longrightarrow\;
    \sum_j T_j \Wdiss^j\om .
\]

\section{Connections}
\label{sec:connections}

In this appendix, we establish the connections between our element-wise thermal-noise formulation and other approaches commonly used in the literature. We first relate our result to Saulson's admittance formulation and show how the thermal-noise contributions of individual elements emerge from combinations of generalized stiffnesses. We then connect our approach to the stored-energy formulation used in continuous systems and develop its representation in terms of normal modes, examining the approximations under which individual modal noise contributions can be treated separately.
\subsection{Connection to Saulson's formulation}
\label{sec:Saulson}

The classic derivation of thermal noise in mechanical systems from the fluctuation--dissipation theorem \cite{FDT} comes from Peter Saulson in 1990 \cite{Saulson1990jc}.
Saulson starts with a force-noise description similar to our \cref{eq:SFj}, and then moves directly to an expression for the readout thermal noise equivalent to
\begin{equation}
    S_{x_R}(\omega)
    =
    \frac{4 k_B T}{\omega^2}
    \real{Y_R(\omega)},
\end{equation}
in our notation, where $Y_R(\omega)$ is the mechanical admittance at the readout.
Using
\begin{equation}
    Y_R(\omega)
    =
    i\omega C_R(\omega),
\end{equation}
where $C_R(\omega)=x(\omega)/f(\omega)$ is the scalar dynamic compliance at the readout, Saulson's expression becomes
\begin{equation}
    S_{x_R}(\omega)
    =
    -\frac{4 k_B T}{\omega}
    \imag{C_R(\omega)}.
    \label{eq:saulson_in_our_notations}
\end{equation}

Introducing the corresponding scalar dynamic stiffness
\begin{equation}
    D_R(\omega)
    =
    \frac{1}{C_R(\omega)},
\end{equation}
we have
\begin{equation}
    \imag{C_R(\omega)}
    =
    -|C_R(\omega)|^2\imag{D_R(\omega)},
\end{equation}
leading to 
\begin{equation}
    S_{x_R}(\omega)
    =
    \frac{4 k_B T}{\omega}|C_R\om|^2
    \imag{D_R(\omega)}.
\end{equation}

The dynamic stiffness $D_R(\omega)$ contains the contributions of all the mechanical components of the system, including elastic elements as well as inertial masses. In the frequency domain, a mass $m$ can be represented in the network description by an inertial element associated with the generalized lossless stiffness
\begin{equation}
    k_m(\omega)=-m\omega^2,
\end{equation}
with one end effectively connected to the inertial frame. The lumped mechanical system can therefore be represented as a network of generalized stiffnesses interconnected in series and parallel. In this representation, the stiffnesses associated with masses are purely real, whereas dissipation is carried by the imaginary parts of the elastic elements. By successively reducing this generalized-stiffness network through series and parallel combinations, the dissipative part of $D_R\om$ can be expressed as a sum of the dissipative contributions of the individual elements, each weighted by the appropriate force-transfer factor. This leads to the element-wise form of \cref{eq:Sx A}, whose general derivation was given in \cref{sec:derivation}.

It is therefore sufficient to consider how the dissipative contributions combine for elements connected locally in parallel and in series, since repeated application of these two elementary reductions extends to all mechanical networks that can be decomposed as series/parallel elements. In addition, these two cases lead to particularly simple expressions for the relative thermal-noise contributions of individual elements.

Consider a set of elements connected in parallel at some location within the mechanical system. We denote by $G_\parallel\om$ the force-transfer factor from this equivalent parallel branch to the readout $x_R$, following the definition in \cref{eq:Gj_def}. We can decompose the equivalent generalized stiffness into its elementary contributions

\begin{equation}
    k_{\parallel}(\omega)
    =
    \sum_{j\in\parallel} k_j(\omega),
\end{equation}
and its imaginary part therefore satisfies

\begin{equation}
    \imag{k_{\parallel}(\omega)}
    =
    \sum_{j\in\parallel}\imag{k_j(\omega)}.
\end{equation}
The contribution of the equivalent spring $k_\parallel\om$ to the thermal noise at the readout $x_R$ becomes
\begin{equation}
    S_{{x_R}\parallel}(\omega)
    =
    \frac{4 k_B T}{\omega}
    |C_R\om|^2|G_{\parallel}(\omega)|^2
    \sum_{j\in\parallel}\imag{k_j(\omega)}
    =
    |C_R\om|^2|G_{\parallel}(\omega)|^2
    \sum_{j\in\parallel}\Sfj(\omega).
\end{equation}

The thermal-force noise associated with the equivalent parallel stiffness is therefore consistently given by
\begin{equation}
    S_f^\parallel(\omega)
    =
    \sum_{j\in\parallel}\Sfj(\omega)
    =
    \frac{4 k_B T}{\omega}
    \imag{k_{\parallel}(\omega)}.
\end{equation}

This shows that the individual thermal-noise sources can be decomposed after deriving the equations of motion of the equivalent system.
This also directly recovers the simple relative contribution of parallel elastic elements given in \cref{eq:parallel thermal noise comparison}.

It may be useful to note that the same equivalent parallel stiffness may be written as
\begin{equation}
    k_\parallel(\omega)
    =
    k'_\parallel(\omega)
    \left[1+i\phi_\parallel(\omega)\right],
\end{equation}
with
\begin{equation}
    k'_\parallel(\omega)
    =
    \sum_{j\in\parallel} k'_j(\omega),
    \qquad
    \phi_\parallel(\omega)
    =
    \frac{1}{k'_\parallel(\omega)}
    \sum_{j\in\parallel} k'_j(\omega)\phi_j(\omega).
\end{equation}

Consider a set of elements connected in series at some location within the mechanical system. We denote by $G_+\om$ the force-transfer factor from this equivalent series branch to the readout $x_R$.
We can decompose the equivalent generalized stiffness into its elementary contributions:
\begin{equation}
    k_{+}(\omega)
    =
    \left(
    \sum_{j\in+}
    \frac{1}{k_j(\omega)}
    \right)^{-1},
\end{equation}
and its imaginary part therefore satisfies

\begin{equation}
    \imag{k_{+}(\omega)}
    =
    \left|
    k_{+}\om
    \right|^2\sum_{j \in +}\frac{\imag{k_j\om}}{|k_j\om|^2}.
\end{equation}
The contribution of the equivalent spring $k_+\om$ to the thermal noise at the readout $x_R$ becomes

\begin{equation}
    S_{x_{R}+}(\omega)
    =
    |C_{R}(\omega)|^2|G_+\om|^2
    \sum_{j\in +}
    |G_j^{(+)}(\omega)|^2
    \Sfj\om,
\end{equation}
where
\begin{equation}
    G_j^{(+)}(\omega)
    =
    \frac{k_+(\omega)}{k_j(\omega)}
    =
    \frac{f_+(\omega)}{f_j(\omega)},
\end{equation}
is the additional force-transfer factor from element $j$ to the equivalent series branch. It follows directly from the usual force division in a series spring network and therefore appears naturally when the equivalent branch is decomposed into its constituent elements.

As in the parallel case, we can define the thermal-force noise of the equivalent spring $k_{+}\om$:

\begin{equation}
S_f^+\om
=
\sum_{j \in +}|G_j^{(+)}\om|^2\Sfj\om
=
\frac{4 k_B T}{\omega}\imag{k_+\om}.
\end{equation}
This also directly recovers the simple relative contribution of two series elastic elements given in \cref{eq:series ratio}.

It may be useful to note that the same equivalent series stiffness may be written as

\begin{equation}
    k_{+}(\omega)
    =
    k'_{+}(\omega)
    \left[1+i\phi_+(\omega)\right],
\end{equation}
which simplifies for light damping $\phi_j\ll1$ for $j\in+$,
\begin{equation}
    k'_{+}(\omega)
    \simeq
    \left(
    \sum_{j\in+}
    \frac{1}{k'_j(\omega)}
    \right)^{-1},
    \qquad
    \phi_+(\omega)
    \simeq
    k'_{+}(\omega)
    \sum_{j\in+}
    \frac{\phi_j(\omega)}{k'_j(\omega)}.
\end{equation}

\subsection{Connection to thermal noise in continuous systems}
\label{app:continuous}
\subsubsection{Connection to Levin's stored-energy formulation}
\label{sec:FEA}

Our approach to computing thermal noise involves approximating mechanical systems as a small number of idealized lumped elements that either have mass or stiffness, but not both.
When a more detailed representation of a system is required, finite-element analysis (FEA) is typically employed to represent extended bodies. 

To compute thermal noise in FEA, the usual approach essentially follows the Levin method \cite{fused2020,3Gsusp};
a generalized force $f_R(\omega)$ is applied to the readout (i.e., as a pressure field with the same form factor that determines the readout variable $x_R$ from the strain field) and the strain-energy distribution of the resulting deformation is computed.

In the lumped-element notation, for each element $j$ with nonzero elastic stiffness $k_j'(\omega)$ and elongation $y_j$, the maximum elastic strain energy in that element when driven is
\begin{equation}
  \Umax^j\om
  =
  \frac12 k'_j\om \asqr{ y_j(\omega)},
 \label{eq:Umax j}
\end{equation}
thus, the power dissipated
in that element, averaged over one cycle, is
\begin{equation}    
    \Wdiss^j\om
    =
    \omega \, \phi_j\om \, \Umax^j\om,
\end{equation}
and total dissipated power in the system is
\begin{equation}    
    \Wdiss\om
    =
    \omega \sum_j \phi_j(\omega) \Umax^j\om.
\end{equation}
Using the temperature-weighted form of Levin's formula described in Appendix~\ref{sec:derivation} gives
\begin{equation}    
    \SxR\om
    =
    \sum_j
    \phi_j\om
    \frac{\Umax^j\om}{|f_R\om|^2}\frac{8 \kBTj}{\omega}.
    \label{eq:levinenergy}
\end{equation}

Substituting $\phi_j=k_j''/k_j'$ and $\Umax^j$, and applying reciprocity as in Appendix~\ref{sec:derivation}, recovers \cref{eq:Sx A}.
The per-element energy can be normalized by the sum of the individual element peaks as
\begin{equation}
\label{eq:Sx FEA A}
    \SxR\om
    =
    \frac{2U\om}{|f_R\om|^2}
    \sum_j
    \frac{4 k_B T_j}{\omega}
    \phi_j\om
    \frac{\Umax^j\om}{U\om},
\end{equation}
where \(U = \sum_j \Umax^j\).

For finite elements, the corresponding elastic energy and dissipated
power are obtained by integrating over each element's volume.
The energy fractions describe the full driven deformation and retain
interference between modal contributions. We next construct the modal
response and examine when separate modal noise contributions can be used.

\subsubsection{Normal-mode decomposition}
\label{sec:normal modes}
The modal representation below assumes a positive-definite mass
matrix (i.e., no massless nodes). Let $\mK(\omega)$ be the complex stiffness matrix describing the
system's restoring and dissipative forces.
The normal-mode vectors $\vpsi_n(\omega)$ are nontrivial solutions
of the generalized eigenvalue problem
\begin{equation}
    \mK\om\,\vpsi_n\om
    = \lambda_n\om\,\mM\,\vpsi_n\om,
    \label{eq:NM eigenvalues}
\end{equation}
where the eigenvalues $\lambda_n(\omega)$ satisfy
\begin{equation}
    \det[\mK\om - \lambda_n\om\mM] = 0,
\end{equation}
for each value of $\omega$.
The frequency-dependent mode-shape vectors $\vpsi_n\om$ are the \emph{right eigenvectors} that solve \cref{eq:NM eigenvalues}
while the \emph{left eigenvectors} $\vpsi_n^L\om$ solve 
\begin{equation}
\vpsi^L_n\om \mK\om - \lambda_n^L\om \, \vpsi^L_n\om \mM  = 0.
\end{equation}
Since $\mK\om$ and $\mM$ are symmetric, transposing
\cref{eq:NM eigenvalues} gives the corresponding left eigenvalue
problem with the same eigenvalue $\lambda_n(\omega)$.
Thus, $\vpsi_n^T(\omega)$ can be chosen as the left eigenvector.

The resulting modal basis is therefore
$\mM$-orthogonal and $\mK(\omega)$-orthogonal. We assume that
they form a complete basis, chosen to be $\mM$-orthogonal also
within any degenerate eigenspaces, and exclude defective
eigenproblems. These orthogonality conditions do not fix the
normalization of the basis vectors.
Since $\vpsi_n$ are generally complex, we normalize with

\begin{equation}
\vpsi_n^\dagger\om \, \vpsi_n\om = 1,    
\end{equation}
while we use overall phase freedom to ensure that the modal mass scalar
\begin{equation}
\mn\om
=
\vpsi_n^T\om \,  \mM \, \vpsi_n\om
\end{equation}
is real and positive.
We similarly define the modal stiffness scalar $\kn$ as
\begin{equation}
\kn\om
=
\vpsi_n^T\om  \, \mK\om \, \vpsi_n\om,
\end{equation}
which may be complex
(i.e., $\kn\om = \kn'\om + i\kn''\om$),
leading to the modal loss angle
$\phin\om = \kn''\om / \kn'\om$.
For completeness we note that
\begin{equation}
    \mt{I}
    =
    \nmsum \vpsi_n\om \frac{\vpsi_n^T\om \, \mM}{\mn\om}.
\end{equation}

The connection between $\lambda_n\om$ and the resonance frequencies of the system is somewhat subtle, since the eigenvalues are found \emph{as a function of $\omega$}.
Rather than dealing with the counterintuitive function $\omega_n\om$ and attempting to analytically continue $\mK\om$ into the complex plane, we identify the system resonances with the frequencies $\omega_n$ for which $\real{\lambda_n\omn} = \omega_n^2$, such that
\begin{equation}
\lambda_n\omn = \omega_n^2 (1 + i \phin\omn),
\end{equation}
and ignore pathological systems where this equality is satisfied by more than one value of $\omega_n$ for a given normal mode $n$.
Since $\mn$ is real, this results in
\begin{equation}
\kn'\omn = \omega_n^2 \, \mn\omn.
\end{equation}

\subsubsection{Modal compliance and transfer factors}
\label{sec:NM compliance}

$\mM$-orthogonality of the eigenvectors implies simple forms for modal dynamic stiffness and compliance
\begin{equation}
\dn\om = \kn\om - \omega^2 \, \mn\om
\quad\text{and}\quad
\cn\om = \frac{1}{\kn\om - \omega^2 \, \mn\om}.
\end{equation}
Since
\begin{equation}
\kn\omn = \omega_n^2 (1 + i \phin\omn) \mn\omn,
\end{equation}
we recover the expected near-resonance response
\begin{equation}
\cn\om \simeq \frac{1}{\kn'\om} H_n\om
\quad\text{where}\quad
H_n\om \equiv \frac{1}{1 + i \phin\om - (\omega / \omega_n)^2},
\label{eq:NM compliance near-resonance}
\end{equation}
if $\kn\om$ and $\bar{m}_n(\omega)$ are at least approximately constant for $\omega \sim \omega_n$.

The displacement vector associated with harmonic motion of normal modes is given by
\begin{equation}
\vx(\omega) = \nmsum a_n(\omega) \vpsi_n\om,
\end{equation}
where $a_n(\omega)$ is the complex amplitude of motion of normal mode $n$
at frequency $\omega$.
Multiplying on the left by $\vpsi_n^T \mM / \mn$, we obtain the modal decomposition of $\vx$ to be
\begin{equation}
a_n(\omega) = \frac{\vpsi_n^T\om \, \mM}{\mn\om} \, \vx\om.
\end{equation}
The modal compliance can also be used to determine the modal amplitude $a_n$ due to a driving force $\vf_A$
\begin{equation}
    a_n\om = \cn\om \, \vpsi_n\om^T \vf_A\om = \cn\om \, \alpha_n^A\om f_A\om 
\quad\text{where}\quad
\alpha_n^A\om  \equiv \vpsi_A^T \vpsi_n\om
\end{equation}
is the coupling coefficient of the generalized force $\vf_A$ into the $n^\text{th}$ normal mode.
The corresponding stretch of element $j$ is given by
\begin{equation}
y_j\om = \ve_j^T \vx\om = \nmsum a_n\om \, \alpha_n^j\om,
\end{equation}
where
$\alpha_n^j\om = \ve_j^T \vpsi_n\om$ is the complex coefficient that maps mode shape to elongation.

We can use this to find the modal element $f_j$-to-$x_A$ compliance
\begin{equation}
\CAj\om \equiv \vpsi_A^T\om \mC\om \ve_j
=
\nmsum \! \CAnj\om
\quad\text{where}\quad
\CAnj\om \equiv \alpha_n^A\om \, \cn\om \, 
\alpha_n^j\om.
\end{equation}
Similarly, the $f_A$-to-$x_A$ compliance is
\begin{equation}
C_A\om = \nmsum \CAn\om
\quad\text{where}\quad
\CAn\om \equiv \alpha_n^A\om^2 \, \cn\om,
\end{equation}
and the $f_j$-to-$f_A$ force transfer factor
\begin{equation}
\GAj\om = \frac{\nmsum \CAnj\om}{\nmsum \CAn\om}.
\end{equation}
This lets us write a modal expression for thermal noise
based on \cref{eq:Sxa_direct}
\begin{equation}
    S_{x_A}(\omega)
    =
    \sum_j
    \Sfj\om \,
    \left|
    \nmsum \CAnj\om
    \right|^2
    =
    \asqr{C_A\om}
    \sum_j
    \Sfj\om \,
    \left|
    \frac{\sum_n \CAnj \om}{\sum_n \CAn \om}
    \right|^2,
\label{eq:NM Sxa general}
\end{equation}
with the rightmost version similar to \cref{eq:Sx B}.

\subsubsection{Modal decomposition of stored energy}
\label{sec:NM energy}

Typically, the use of normal-mode decompositions in thermal-noise calculations involves an apparently direct application of Levin's approach in \cref{eq:levin_omega}.
This method involves estimating the dissipation in each element by the fraction of elastic strain energy stored.
Doing this correctly turns out to be somewhat subtle due to differences in the modal and elemental energy decomposition, as shown below.

We can use \cref{eq:Umax j} to arrive at an expression for the energy stored in element $j$ 
\begin{equation}
\Umax^j\om = \frac12 k'_j\om \asqr{ \nmsum a_n\om \, \alpha_n^j\om }, 
\end{equation}
and the sum of the individual element peaks
\begin{equation}
U\om = \frac12 \sum_j k'_j\om \asqr{ \nmsum a_n\om \, \alpha_n^j\om }, 
\end{equation}
with the power dissipated, averaged over one cycle in element $j$ given by
\begin{equation}
\Wdiss^j\om = \omega  \, \phi_j\om \, \Umax^j\om.
\end{equation}
If $a_n$ is the result of a driving force $\vf_A$, the peak energy in element $j$ is 
\begin{equation}
\UmaxA^j\om = \frac12 k'_j\om \asqr{ \nmsum \alpha_n^A\om \, \cn\om\, \alpha_n^j\om \, f_A\om} = \frac12 k'_j\om \asqr{\CAj\om \, f_A\om},
\end{equation}
With $U_A=\sum_j\UmaxA^j$, the thermal-noise expression is
\begin{align}
    S_{x_A}(\omega)
    & =
    \frac{U_A\om}{\asqr{f_A\om}}
    \sum_j
    \frac{\Sfj\om}{\frac12 k_j'\om} \,
    \frac{\UmaxA^j\om}{U_A\om} \\
    & =
    2\frac{U_A\om}{\asqr{f_A\om}}
    \sum_j
    \frac{4 \kBTj}{\omega} \,
    \phi_j\om \,
    \frac{\UmaxA^j\om}{U_A\om}.
\label{eq:NM Sxa general Umax}
\end{align}
The leading coefficient in \cref{eq:NM Sxa general Umax} can be decomposed into modal response functions as
\begin{align}
    2\frac{U_A\om}{\asqr{f_A\om}}
    &=
    \sum_j k_j'\om \,
    \vpsi_A^T \mC\om
    \ve_j
    \ve_j^T \mC\om^*
    \vpsi_A^*
    \\
    &=
    \vpsi_A^T \mC\om \,
    \mK'\om \,
    \mC\om^*
    \vpsi_A^*
    \\
    &=
    \sum_{n,m}^{\rm modes}
    \left[
    \alpha_n^A\om \,
    \cn\om
    \right] \,
    \left[
    \vpsi_n^T\om \mK'\om \vpsi_m^*\om
    \right]
    \,
    \left[\modal{c}_m\om \,\alpha_m^A\om\right]^*
    \\
    &=
    \sum_{n,m}^{\rm modes}
    \Gamma_{nm}\om
    \sqrt{\kn'\om \modal{k}_m'\om}
    \left[
    \alpha_n^A\om \cn\om
    \right]
    \left[
    \alpha_m^A\om \modal{c}_m\om
    \right]^*,
\end{align}
where the dimensionless modal energy-overlap matrix is
\begin{equation}
    \Gamma_{nm}\om
    \equiv
    \frac{
    \vpsi_n^T\om \mK'\om \vpsi_m^*\om
    }{
    \sqrt{\modal{k}_n'\om \modal{k}_m'\om}
    }
    =
    \frac{
    \sum_j k_j' \alpha_n^j \, \alpha_m^{j\,*}
    }{
    \sqrt{\modal{k}_n'\om \modal{k}_m'\om}
    }.
\end{equation}

In the following subsections, we consider cases that allow approximations of $\mGamma$ and simplifications of $\SxA$.

\subsubsection{Low-loss limit}
\label{sec:NM low loss limit}

In the zero-loss limit, the complex mode shapes reduce to the real conservative
modes, $\vpsi_n\rightarrow \vc{u}_n$, for which
$\vc{u}_n^T\mK'\vc{u}_m = \kn \delta_{nm}$. Therefore
\begin{equation}
    \Gamma_{nm}
    =
    \frac{\vpsi_n^T\mK'\vpsi_m^*}
    {\sqrt{\kn'\bar{k}_m'}}
    \rightarrow
    \frac{\vc{u}_n^T\mK'\vc{u}_m}
    {\sqrt{\kn \modal{k}_m}}
    =
    \delta_{nm},
\end{equation}
as $\phi_j\rightarrow0$.
Thus, the off-diagonal elements of
$\mGamma$ measure the failure of lossy mode shapes to remain orthogonal with respect to the conservative stored-energy metric.
In fact, for low-loss nondegenerate modes 
\begin{equation}
    \Gamma_{nn} = 1 + O(\phi^2)
    \quad\text{and}\quad
    \Gamma_{nm} = O(\phi)
    \quad (n\neq m),
\end{equation}
in general.

If the modal energy-overlap matrix is approximately diagonal,
\begin{equation}
    \Gamma_{nm}\om \simeq \gamma_n\om \delta_{nm},
\end{equation}
then
\begin{align}
    \gamma_n\om
    \equiv
    \mGamma_{nn}\om
    &=
    \frac{\vpsi_n^T\om \mK'\om \vpsi_n^*\om}
    {\kn'\om} \\
    &=
    \frac{\vpsi_n^T\om \real{\mK\om} \vpsi_n^*\om}
    {\real{\vpsi_n^T\om \, \mK\om \, \vpsi_n\om}} \\
    &=
    \frac{\sum_j k_j'\om \asqr{\alpha_n^j\om}}
    {\sum_j \real{ k_j\om \, (\anj\om)^2}},
\end{align}
with $\gamma_n \rightarrow 1$ for real $\anj$ (e.g., in conservative systems).
For approximately diagonal $\mGamma$, the leading coefficient in
\cref{eq:NM Sxa general Umax} reduces to
\begin{equation}
    2\frac{U_A\om}{\asqr{f_A\om}}
    \simeq
    \nmsum
    \gamma_n\om
    \modal{k}_n'\om
    \asqr{\alpha_n^A\om \cn\om}.
\end{equation}
For approximately real mode shapes, $\gamma_n\simeq1$, and
using $\cn\om\simeq H_n\om/\kn'\om$, the noise spectrum becomes
\begin{equation}
    S_{x_A}(\omega)
    \simeq
    \left(
        \nmsum
        \frac{\asqr{\alpha_n^A\om H_n\om}}{\kn'\om}
    \right)
    \left(
        \sum_j
        \frac{4\kBTj}{\omega}\,
        \phi_j\om\,
        \frac{\UmaxA^j\om}{U_A\om}
    \right).
    \label{eq:NM Sxa low-loss limit}
\end{equation}
For $A=R$, this gives the readout spectrum with the full driven energy fractions. These fractions retain interference between modal contributions within each element.

\subsubsection{Conditions for separating modal noise contributions}
In practice, several additional approximations are sometimes made
that go beyond those required to arrive at \cref{eq:NM Sxa low-loss limit}.
Returning to \cref{eq:levinenergy}, one may attempt to
break down the thermal noise into contributions from each normal
mode, such that
\begin{equation}
    \SxR(\omega)
     \simeq
    \sum_j
    \frac{8 \kBTj}{\omega} \,
    \phi_j\om \,
    \left(
    \nmsum
    \frac{\Umaxn^j\om}{|f_R\om|^2}
    \right).
\label{eq:Sx FEA C}
\end{equation}
A sufficient condition for the approximation in \cref{eq:Sx FEA C} is that the elongation $y_j$ of each dissipative element, defined in \cref{eq:Umax j}
\begin{equation}
y_j\om = \nmsum a_n\om \, \alpha_n^j 
\quad\text{where}\quad
\alpha_n^j = \ve_j^T \vpsi_n,
\end{equation}
be dominated, at each frequency, by a single mode, giving
\begin{equation}
\Umax^j\om
=
\frac12 k'_j\om \asqr{\nmsum a_n\om \, \alpha_n^j\om}
\simeq
\frac12 k'_j\om \nmsum \asqr{a_n\om \, \alpha_n^j\om}.
\end{equation}
Here, cross terms between modal contributions to the element's
energy are negligible. The full modal sum is retained because
the dominant mode may differ between elements and change with
frequency. This allows
\begin{equation}
\Umax^j\om
\simeq
\nmsum \Umaxn^j\om
\quad\text{with}\quad
\Umaxn^j\om
\equiv
\frac12 k'_j\om \asqr{ a_n\om \, \alpha_n^j\om }.
\end{equation}
Finally, using this approximation and $a_n\om = \modal{c}_n \alpha_n^R f_R\om$
we obtain
\begin{equation}
    \SxR(\omega)
     \simeq
    \sum_j
    \frac{4 \kBTj}{\omega} \,
    k_j''\om \,
    \left(
    \nmsum
    \asqr{
    \frac{\alpha_n^R \, \alpha_n^j \, H_n\om}{\kn'\om}}
    \right).
\label{eq:Sx FEA D}
\end{equation}

This approximation is appropriate near resonances where one mode
dominates, but may not be valid far from resonance and should
therefore be used with care.

\bibliography{bibliography}

@article{Nguyen2015,
  author  = {Nguyen, Thanh T-H. and Mow-Lowry, Conor M. and
             Slagmolen, Bram J. J. and Miller, John and
             Mullavey, Adam J. and Go{\ss}ler, Stefan and
             Altin, Paul A. and Shaddock, Daniel A. and
             McClelland, David E.},
  title   = {Frequency dependence of thermal noise in gram-scale cantilever flexures},
  journal = {Phys. Rev. D},
  volume  = {92},
  number  = {11},
  pages   = {112004},
  year    = {2015},
  doi     = {10.1103/PhysRevD.92.112004}
}

@article{Amos2024,
  author  = {Amos, C. C. and Prasad, M. and Cannon, K. M. and Dreyer, C. B.},
  title   = {Velocity Measurements of Powdered Rock at Low Confining
             Pressures and Comparison to Lunar Shallow Seismic Velocity},
  journal = {J. Geophys. Res. Planets},
  year    = {2024},
  volume  = {129},
  number  = {5},
  pages   = {e2024JE008287},
  doi     = {10.1029/2024JE008287}
}

@article{Williams2017,
  author  = {Williams, Jean-Pierre and Paige, David A. and
             Greenhagen, Benjamin T. and Sefton-Nash, Elliot},
  title   = {The global surface temperatures of the {Moon} as measured
             by the {Diviner Lunar Radiometer Experiment}},
  journal = {Icarus},
  volume  = {283},
  pages   = {300--325},
  year    = {2017},
  doi     = {10.1016/j.icarus.2016.08.012}
}

@article{Saffarieh2026,
  author  = {Saffarieh, Pooya and Holland, Nathan A. and
             Valentini, Michele and van Dongen, Jesse and
             Mitchell, Alexandra and Sijtsma, Sander and
             Numic, Armin and Hakvoort, Wouter and
             Mow-Lowry, Conor},
  title   = {Multi-scale optimal control for {Einstein Telescope}
             active seismic isolation},
  journal = {Class. Quantum Grav.},
  volume  = {43},
  pages   = {025004},
  year    = {2026},
  doi     = {10.1088/1361-6382/ae3422}
}

@article{Aston2012,
  author  = {Aston, S M and
             Barton, M A and
             Bell, A S and
             Beveridge, N and
             Bland, B and
             Brummitt, A J and
             Cagnoli, G and
             Cantley, C A and
             Carbone, L and
             Cumming, A V and
             Cunningham, L and
             Cutler, R M and
             Greenhalgh, R J S and
             Hammond, G D and
             Haughian, K and
             Hayler, T M and
             Heptonstall, A and
             Heefner, J and
             Hoyland, D and
             Hough, J and
             others},
  title   = {Update on quadruple suspension design for {Advanced LIGO}},
  journal = {Class. Quantum Grav.},
  volume  = {29},
  number  = {23},
  pages   = {235004},
  year    = {2012},
  doi     = {10.1088/0264-9381/29/23/235004}
}

@article{Cagnoli1999Wires,
  author  = {Cagnoli, G. and Gammaitoni, L. and Kovalik, J.
             and Marchesoni, F. and Punturo, M.},
  title   = {Low-frequency internal friction in clamped-free thin wires},
  journal = {Phys. Lett. A},
  volume  = {255},
  number  = {4--6},
  pages   = {230--235},
  year    = {1999},
  doi     = {10.1016/S0375-9601(99)00184-X}
}

@misc{Trippe2026,
  author        = {Trippe, James and Polidan, Ronald and
                   Creighton, Teviet and Lognonn{\'e}, Philippe and
                   Panning, Mark and Quetschke, Volker and
                   Izquierdo, Kris and Shapiro, Brett and Jani, Karan},
  title         = {Site selection for the {Laser Interferometer
                   Lunar Antenna ({LILA})}},
  year          = {2026},
  eprint        = {2604.02405},
  archivePrefix = {arXiv},
  primaryClass  = {astro-ph.IM},
  doi           = {10.48550/arXiv.2604.02405}
}

@misc{CEHS,
      title={A Horizon Study for {Cosmic Explorer}: Science, Observatories, and Community}, 
      author={Matthew Evans and Rana X Adhikari and Chaitanya Afle and Stefan W. Ballmer and Sylvia Biscoveanu and Ssohrab Borhanian and Duncan A. Brown and Yanbei Chen and Robert Eisenstein and Alexandra Gruson and Anuradha Gupta and Evan D. Hall and Rachael Huxford and Brittany Kamai and Rahul Kashyap and Jeff S. Kissel and Kevin Kuns and Philippe Landry and Amber Lenon and Geoffrey Lovelace and others},
      year={2021},
      eprint={2109.09882},
      archivePrefix={arXiv},
      primaryClass={astro-ph.IM},
  note = {{Cosmic Explorer} technical report CE-P2100003},
  url = {https://dcc.cosmicexplorer.org/CE-P2100003/public}
}

@article{Smetana2025,
  author  = {Smetana, Jiri and
             Ubhi, Amit Singh and
             Chick, Emilia and
             Prokhorov, Leonid and
             Bryant, John and
             Dmitriev, Artemiy and
             Gill, Alex and
             Koponen, Lari and
             Martynov, Denis and
             Miao, Haixing and
             Cumming, Alan V. and
             Hammond, Giles and
             Frolov, Valery and
             Mittleman, Richard and
             Fritchel, Peter},
  title   = {Sensitivity and control of a six-axis fused-silica seismometer},
  journal = {Phys. Rev. Appl.},
  volume  = {23},
  number  = {2},
  pages   = {024013},
  year    = {2025},
  doi     = {10.1103/PhysRevApplied.23.024013},
  eprint = {2405.13475},
  archivePrefix = {arXiv},
  primaryClass = {physics.ins-det}
}

@article{Gras2017,
  author  = {Gras, S. and Yu, H. and Yam, W. and Martynov, D. and Evans, M.},
  title   = {Audio-band coating thermal noise measurement for {Advanced LIGO} with a multimode optical resonator},
  journal = {Phys. Rev. D},
  volume  = {95},
  number  = {2},
  pages   = {022001},
  year    = {2017},
  doi     = {10.1103/PhysRevD.95.022001}
}

@article{fused2020,
	doi = {10.1088/1361-6382/abac42},
	journal = {Class. Quantum Grav.},
	url = {https://doi.org/10.1088/1361-6382/abac42},
	year = 2020,
	month = {sep},
	volume = {37},
	number = {19},
	pages = {195019},
	author = {A V Cumming and B Sorazu and E Daw and G D Hammond and J Hough and R Jones and I W Martin and S Rowan and K A Strain and D Williams},
	title = {Lowest observed surface and weld losses in fused silica fibres for gravitational wave detectors},
}

@article{3Gsusp,
  author = {Cumming, A. V. and Jones, R. and Hammond, G. D. and Hough, J. and Martin, I. W. and Rowan, S.},
  title = {Large-scale monolithic fused-silica mirror suspension for third-generation gravitational-wave detectors},
  journal = {Phys. Rev. Appl.}, volume = {17}, number = {2}, pages = {024044}, year = {2022},
  doi = {10.1103/PhysRevApplied.17.024044}
}

@article{Caves1980a,
author = {Caves, Carlton M},
journal = {Phys. Rev. Lett.},
pages = {75},
title = {{Quantum-Mechanical Radiation-Pressure Fluctuations in an Interferometer}},
url ={http://journals.aps.org/prl/abstract/10.1103/PhysRevLett.45.75},
volume = {45},
year = {1980},
  doi = {10.1103/PhysRevLett.45.75}
}

@article{instrO3paper,
  title={Sensitivity and performance of the {Advanced LIGO} detectors in the third observing run},
  author={Buikema, A. and Cahillane, C. and Mansell, G. L. and Blair, C. D. and Abbott, R. and Adams, C. and Adhikari, R. X. and Ananyeva, A. and Appert, S. and Arai, K. and Areeda, J. S. and Asali, Y. and Aston, S. M. and Austin, C. and Baer, A. M. and Ball, M. and Ballmer, S. W. and Banagiri, S. and Barker, D. and Barsotti, L. and others},
  journal={Phys. Rev. D},
  volume={102},
  number={6},
  pages={062003},
  year={2020},
  doi = {10.1103/PhysRevD.102.062003},
  eprint = {2008.01301},
  archivePrefix = {arXiv},
  primaryClass = {astro-ph.IM}
}

@ARTICLE{Gretarsson2000,
       author = {{Gretarsson}, Andri M. and {Harry}, Gregory M. and {Penn}, Steven D. and {Saulson}, Peter R. and {Startin}, William J. and {Rowan}, Sheila and {Cagnoli}, Gianpietro and {Hough}, Jim},
        title = "{Pendulum mode thermal noise in advanced interferometers: a comparison of fused silica fibers and ribbons in the presence of surface loss}",
      journal = {Phys. Lett. A},
         year = 2000,
        month = may,
       volume = {270},
       number = {3-4},
        pages = {108-114},
          doi = {10.1016/S0375-9601(00)00295-4},
archivePrefix = {arXiv},
       eprint = {gr-qc/9912057},
 primaryClass = {gr-qc},
}

@article{Saulson1990jc, 
  year    = {1990}, 
  title   = {Thermal noise in mechanical experiments}, 
  author  = {Saulson, Peter R}, 
  journal = {Phys. Rev. D}, 
  doi     = {10.1103/physrevd.42.2437}, 
  pages   = {2437 -- 2445}, 
  volume  = {42}, 

}

@article{levin1998, 
  year    = {1998}, 
  title   = {Internal thermal noise in the {LIGO} test masses: A direct approach}, 
  author  = {Levin, Y}, 
  journal = {Phys. Rev. D}, 
  doi     = {10.1103/physrevd.57.659}, 
  url     = {http://link.aps.org/doi/10.1103/PhysRevD.57.659}, 
  pages   = {659 -- 663}, 
  volume  = {57}, 

}

@article{FDT, 
  year    = {1952}, 
  title   = {On a Theorem of Irreversible Thermodynamics}, 
  author  = {Callen, Herbert and Greene, Richard}, 
  journal = {Phys. Rev.}, 
  doi     = {10.1103/physrev.86.702}, 
  url     = {http://link.aps.org/doi/10.1103/PhysRev.86.702}, 
  pages   = {702 -- 710}, 
  volume  = {86}, 
  month   = {06}
}

@book{textbook:Inman1994,
  title={Engineering vibration},
  author={Inman, Daniel J and Singh, Ramesh Chandra},
  year={1994},
  publisher={Prentice Hall},
  address = {Englewood Cliffs, NJ}
}

@book{textbook:Meirovitch1997,
  title={Principles and techniques of vibrations},
  author={Meirovitch, Leonard},
  publisher={Prentice Hall},
  year={1997},
  address = {Upper Saddle River, NJ}
}

@book{textbook:Cook2007FEA,
  title={Concepts and applications of finite element analysis},
  author={Cook, Robert D. and Malkus, David S. and Plesha, Michael E. and Witt, Robert J.},
  year={2002},
  publisher={John Wiley \& Sons},
  edition = {4th},
  address = {New York}
}

@book{Hunt1954,
  author    = {Hunt, Frederick V.},
  title     = {Electroacoustics: The Analysis of Transduction, and Its Historical Background},
  publisher = {Harvard University Press},
  address   = {Cambridge, Massachusetts},
  year      = {1954}
}

@article{Nyquist1928,
  author  = {Nyquist, Harry},
  title   = {Thermal Agitation of Electric Charge in Conductors},
  journal = {Phys. Rev.},
  volume  = {32},
  number  = {1},
  pages   = {110--113},
  year    = {1928},
  doi     = {10.1103/PhysRev.32.110}
}

@article{Gonzalez1994, 
  year    = {1994}, 
  title   = {Brownian motion of a mass suspended by an anelastic wire}, 
  author  = {Gonzalez, G and Saulson, P R}, 
  journal = {J. Acoust. Soc. Am.}, 
  doi     = {10.1121/1.410467}, 
  url     = {http://scitation.aip.org/content/asa/journal/jasa/96/1/10.1121/1.410467}, 
  pages   = {207 -- 212}, 
  volume  = {96}, 

}

@article{Huang1998,
  author  = {Huang, Y. L. and Saulson, Peter R.},
  title   = {Dissipation Mechanisms in Pendulums and Their Implications for Gravitational Wave Interferometers},
  journal = {Rev. Sci. Instrum.},
  volume  = {69},
  number  = {2},
  pages   = {544--553},
  year    = {1998},
  doi     = {10.1063/1.1148692}
}

@article{Djuric2000,
  author  = {Djuri{\'c}, Zoran},
  title   = {Mechanisms of Noise Sources in Microelectromechanical Systems},
  journal = {Microelectron. Reliab.},
  volume  = {40},
  number  = {6},
  pages   = {919--932},
  year    = {2000},
  month   = jun,
  doi     = {10.1016/S0026-2714(00)00004-4},
  url     = {https://doi.org/10.1016/S0026-2714(00)00004-4}
}

@article{ButtJaschke1995,
  author  = {Butt, Hans-J{\"u}rgen and Jaschke, Manfred},
  title   = {Calculation of Thermal Noise in Atomic Force Microscopy},
  journal = {Nanotechnology},
  volume  = {6},
  number  = {1},
  pages   = {1--7},
  year    = {1995},
  doi     = {10.1088/0957-4484/6/1/001},
  url     = {https://doi.org/10.1088/0957-4484/6/1/001}
}

@article{Bellon2008,
  author  = {Bellon, Ludovic},
  title   = {Thermal Noise of Microcantilevers in Viscous Fluids},
  journal = {J. Appl. Phys.},
  volume  = {104},
  number  = {10},
  pages   = {104906},
  year    = {2008},
  doi     = {10.1063/1.3021102},
  eprint  = {0811.3429},
  archivePrefix = {arXiv},
  url     = {https://doi.org/10.1063/1.3021102}
}

@article{BraginskyLevinVyatchanin1999,
  author  = {Braginsky, V. B. and Levin, Yu. and Vyatchanin, S. P.},
  title   = {How to reduce suspension thermal noise in {LIGO} without improving the {\ensuremath{Q}} of the pendulum and violin modes},
  journal = {Meas. Sci. Technol.},
  volume  = {10},
  number  = {7},
  pages   = {598--606},
  year    = {1999},
  doi     = {10.1088/0957-0233/10/7/305},
  eprint  = {gr-qc/9805031},
  archivePrefix = {arXiv},
  url     = {https://doi.org/10.1088/0957-0233/10/7/305}
}

@article{Harry2002,
  author  = {Harry, Gregory M. and Gretarsson, Andri M. and Saulson, Peter R. and Kittelberger, Scott E. and Penn, Steven D. and Startin, William J. and Rowan, Sheila and Fejer, Martin M. and Crooks, David R. M. and Cagnoli, Gianpietro and Hough, Jim and Nakagawa, Norio},
  title   = {Thermal Noise in Interferometric Gravitational Wave Detectors Due to Dielectric Optical Coatings},
  journal = {Class. Quantum Grav.},
  volume  = {19},
  number  = {5},
  pages   = {897--917},
  year    = {2002},
  doi     = {10.1088/0264-9381/19/5/305},
  eprint  = {gr-qc/0109073},
  archivePrefix = {arXiv},
  url     = {https://doi.org/10.1088/0264-9381/19/5/305}
}

@article{Evans2008,
  title = {Thermo-optic noise in coated mirrors for high-precision optical measurements},
  author = {Evans, M. and Ballmer, S. and Fejer, M. and Fritschel, P. and Harry, G. and Ogin, G.},
  journal = {Phys. Rev. D},
  volume = {78},
  issue = {10},
  pages = {102003},
  numpages = {10},
  year = {2008},
  month = {Nov},
  doi = {10.1103/PhysRevD.78.102003},
  url = {https://link.aps.org/doi/10.1103/PhysRevD.78.102003}
}

@article{Hong2013,
  title = {Brownian thermal noise in multilayer coated mirrors},
  author = {Hong, Ting and Yang, Huan and Gustafson, Eric K. and Adhikari, Rana X. and Chen, Yanbei},
  journal = {Phys. Rev. D},
  volume = {87},
  issue = {8},
  pages = {082001},
  numpages = {27},
  year = {2013},
  month = {Apr},
  doi = {10.1103/PhysRevD.87.082001},
  url = {https://link.aps.org/doi/10.1103/PhysRevD.87.082001}
}

@article{Nakagawa2002,
  author  = {Nakagawa, N. and Gretarsson, A. M. and Gustafson, E. K. and Fejer, M. M.},
  title   = {Thermal Noise in Half-Infinite Mirrors with Nonuniform Loss: A Slab of Excess Loss in a Half-Infinite Mirror},
  journal = {Phys. Rev. D},
  volume  = {65},
  pages   = {102001},
  year    = {2002},
  doi     = {10.1103/PhysRevD.65.102001},
  eprint  = {gr-qc/0105046},
  archivePrefix = {arXiv},
  url     = {https://doi.org/10.1103/PhysRevD.65.102001}
}

@article{Grass2026,
  author  = {Grass, Daniel and Vermeulen, Sander M. and MacMillan, Ian A. O. and McCuller, Lee},
  title   = {High-frequency thermal noise in {Michelson} interferometers},
  year    = {2026},
  eprint  = {2605.16710},
  archivePrefix = {arXiv},
  primaryClass = {physics.ins-det},
  journal = {Phys. Rev. D},
  note = {accepted for publication},
  doi = {10.1103/7kw1-sh6m}
}

@article{Ohishi2000,
  author  = {Ohishi, Naoko and Otsuka, Shigemi and Kawabe, Keita and Tsubono, Kimio},
  title   = {Estimation of Thermal Noise by a Direct Measurement of the Mechanical Conductance},
  journal = {Phys. Lett. A},
  volume  = {266},
  number  = {4--6},
  pages   = {228--233},
  year    = {2000},
  doi     = {10.1016/S0375-9601(99)00867-1},
  url     = {https://doi.org/10.1016/S0375-9601(99)00867-1}
}

@article{Ruggi2025,
  author  = {Ruggi, Paolo and Pinto, Manuel and Trozzo, Lucia and Cella, Giancarlo and Majorana, Ettore and Losurdo, Giovanni and Chessa, Piero and Longo, Alessandro and Vicer{\'e}, Andrea},
  title   = {Mechanical Simulation Tool Based on Impedance Matrices},
  journal = {Phys. Rev. D},
  volume  = {112},
  number  = {2},
  pages   = {022002},
  year    = {2025},
  doi     = {10.1103/PhysRevD.112.022002},
  url     = {https://doi.org/10.1103/PhysRevD.112.022002}
}

@article{Numata2004,
  author    = {Numata, Kenji and Kemery, Amy and Camp, Jordan},
  title     = {Thermal-Noise Limit in the Frequency Stabilization of Lasers with Rigid Cavities},
  journal   = {Phys. Rev. Lett.},
  volume    = {93},
  number    = {25},
  pages     = {250602},
  year      = {2004},
  month     = {Dec},
  doi       = {10.1103/PhysRevLett.93.250602},
  url       = {https://doi.org/10.1103/PhysRevLett.93.250602}
}

@article{HarmsMowLowry2018,
  author  = {Harms, Jan and Mow-Lowry, Conor M.},
  title   = {Suspension-Thermal Noise in Spring--Antispring Systems for Future Gravitational-Wave Detectors},
  journal = {Class. Quantum Grav.},
  volume  = {35},
  number  = {2},
  pages   = {025008},
  year    = {2018},
  doi     = {10.1088/1361-6382/aa9e28},
  eprint  = {1707.07309},
  archivePrefix = {arXiv},
  primaryClass = {gr-qc},
  url     = {https://doi.org/10.1088/1361-6382/aa9e28}
}

@article{Gras2015AMD,
  author  = {Gras, S. and Fritschel, P. and Barsotti, L. and Evans, M.},
  title   = {Resonant dampers for parametric instabilities in gravitational wave detectors},
  journal = {Phys. Rev. D},
  volume  = {92},
  number  = {8},
  pages   = {082001},
  year    = {2015},
  doi     = {10.1103/PhysRevD.92.082001}
}

@article{Biscans2019AMD,
  author  = {Biscans, S. and Gras, S. and Blair, C. D. and Driggers, J. and
             Evans, M. and Fritschel, P. and Hardwick, T. and Mansell, G.},
  title   = {Suppressing parametric instabilities in {LIGO} using low-noise acoustic mode dampers},
  journal = {Phys. Rev. D},
  volume  = {100},
  number  = {12},
  pages   = {122003},
  year    = {2019},
  doi     = {10.1103/PhysRevD.100.122003}
}

@misc{Creighton2025LILA,
  author        = {Creighton, Teviet and Lognonn{\'e}, Philippe and Panning, Mark P. and Trippe, James and Quetschke, Volker and Jani, Karan},
  title         = {Fundamental Noise and Gravitational-Wave Sensitivity of the {Laser Interferometer Lunar Antenna} ({LILA})},
  year          = {2025},
  url           = {https://arxiv.org/abs/2508.18437},
  eprint        = {2508.18437},
  archivePrefix = {arXiv},
  primaryClass  = {gr-qc}
}

@article{Mitchell1972,
  author  = {Mitchell, James K. and Bromwell, Leslie G. and Carrier, III, W. David and Costes, Nicholas C. and Scott, Ronald F.},
  title   = {Soil Mechanical Properties at the {Apollo 14} Site},
  journal = {J. Geophys. Res.},
  volume  = {77},
  number  = {29},
  pages   = {5641--5664},
  year    = {1972},
  doi     = {10.1029/JB077i029p05641}
}

@article{Gazetas1991,
  author  = {Gazetas, George},
  title   = {Formulas and Charts for Impedances of Surface and Embedded Foundations},
  journal = {J. Geotech. Eng.},
  volume  = {117},
  number  = {9},
  pages   = {1363--1381},
  year    = {1991},
  doi     = {10.1061/(ASCE)0733-9410(1991)117:9(1363)}
}

@article{Penn2006FusedSilica,
  author  = {Penn, Steven D. and Ageev, Alexander and Busby, Dan and Harry, Gregory M. and Gretarsson, Andri M. and Numata, Kenji and Willems, Phil},
  title   = {Frequency and surface dependence of the mechanical loss in fused silica},
  journal = {Phys. Lett. A},
  volume  = {352},
  number  = {1--2},
  pages   = {3--6},
  year    = {2006},
  doi     = {10.1016/j.physleta.2005.11.046}
}

@article{Onodera2021LunarQ,
  author  = {Onodera, K. and
             Kawamura, T. and
             Tanaka, S. and
             Ishihara, Y. and
             Maeda, T.},
  title   = {Numerical Simulation of Lunar Seismic Wave Propagation:
             Investigation of Subsurface Scattering Properties Near
             {Apollo 12} Landing Site},
  journal = {J. Geophys. Res. Planets},
  year    = {2021},
  volume  = {126},
  pages   = {e2020JE006406},
  doi     = {10.1029/2020JE006406}
}

@article{Gonzalez2000,
  author  = {Gonz{\'a}lez, Gabriela},
  title   = {Suspensions Thermal Noise in the {LIGO} Gravitational Wave Detector},
  journal = {Class. Quantum Grav.},
  volume  = {17},
  number  = {21},
  pages   = {4409--4435},
  year    = {2000},
  doi     = {10.1088/0264-9381/17/21/305},
  eprint  = {gr-qc/0006053},
  archivePrefix = {arXiv}
}

@article{Puppo2010,
  author  = {Puppo, Paola},
  title   = {A Thermal Noise Model for a Branched System of Harmonic Oscillators},
  journal = {J. Phys.: Conf. Ser.},
  volume  = {228},
  pages   = {012031},
  year    = {2010},
  doi     = {10.1088/1742-6596/228/1/012031}
}

@article{Komori2018,
  title = {Direct approach for the fluctuation-dissipation theorem under nonequilibrium steady-state conditions},
  author = {Komori, Kentaro and Enomoto, Yutaro and Takeda, Hiroki and Michimura, Yuta and Somiya, Kentaro and Ando, Masaki and Ballmer, Stefan W.},
  journal = {Phys. Rev. D},
  volume = {97},
  number = {10},
  pages = {102001},
  year = {2018},
  doi = {10.1103/PhysRevD.97.102001},
  eprint = {1803.00585},
  archivePrefix = {arXiv},
  primaryClass = {gr-qc}
}

\end{document}